\documentclass[%
notitlepage,
reprint,
superscriptaddress,
amsmath,amssymb,
aps,
]{revtex4-2}

\usepackage{xcolor}
\usepackage{lipsum}
\usepackage{graphicx}
\usepackage{dcolumn}
\usepackage{bm}
\usepackage{hyperref}

\usepackage{epstopdf}

\RequirePackage[normalem]{ulem}
\RequirePackage{color}

\newcommand{\bra}[1]{\langle #1 \vert}
\newcommand{\ket}[1]{\vert #1 \rangle}

\usepackage{ulem}

\begin{document}
	
	
\title{Discrete-time quantum walk dispersion control through long-range correlations}
	
\affiliation{Laborat\' orio de F\'isica Te\' orica e Computacional, Departamento de F\'isica, Universidade Federal de Pernambuco, 50670-901 Recife, PE, Brazil}
%
%
\affiliation{Grupo de Física Computacional Aplicada, Instituto Federal de Alagoas, 57020-600, Macei\' o, Alagoas, Brazil}
\affiliation{Instituto de F\'isica, Universidade Federal de Alagoas, 57072-900 Macei\'o, Alagoas, Brazil}

\author{A. R. C. Buarque$^{1,2}$}
	
\email{abuarque@professor.educ.al.gov.br}
\author{F. S. Passos$^{2}$}
\author{W. S. Dias$^{3}$}
\author{E. P. Raposo$^{1}$}

\begin{abstract}
We investigate the evolution dynamics of inhomogeneous discrete-time one-dimensional quantum walks displaying long-range correlations in both space and time. 
The associated quantum coin operators are built to exhibit a random inhomogeneity  distribution of long-range correlations embedded in the time evolution protocol through a fractional Brownian motion with spectrum following a power-law behavior, $S(k) \sim 1/k^\nu$. 
The power-law correlated disorder encoded in the phases of the quantum coin is shown to give rise to a wide variety of spreading patterns of the qubit states, from localized to subdiffusive, diffusive, and superdiffusive (including ballistic) behavior, depending on the relative strength of the parameters driving the correlation degree. 
Dispersion control is then possible in one-dimensional discrete-time quantum walks by suitably tunning the long-range correlation properties assigned to the inhomogeneous quantum coin operator.
\end{abstract}


\maketitle
	
\section{Introduction}
	
Discrete-time quantum walks (DTQWs)~\cite{n1,n2,venegas2012quantum,kempe2003quantum} have become a central topic of research in recent years in part due to the characteristic of faster propagation over time when compared to classical random walks. 
Originally introduced by Ahanorov, Davidovich, and Zagury~\cite{PhysRevA.48.1687} as a generalization of the standard (classical) random walks, DTQWs readily drew considerable attention for exhibiting anomalous dynamics with dispersion increasing linearly (i.e., ballistically) with time, in contrast with the square-root time dynamics of their classical (Brownian) counterpart.

Over the last three decades, DTQWs have found numerous applications~\cite{n1,n2,venegas2012quantum,kempe2003quantum} in emerging quantum technologies and complex systems, across various fields, including physics, mathematics, engineering, computer science, and biology. 
In particular, in physics DTQWs have been utilized in simulations of complex physical systems and for investigating diverse relevant topics such as quantum computation~\cite{PhysRevA.81.042330,singh2021universal}, quantum algorithms~\cite{PhysRevA.67.052307,portugal2013quantum}, quantum entanglement~\cite{PhysRevA.85.022322}, cybersecurity~\cite{abd2021quantum,abd2020providing}, strongly correlated phenomena~\cite{PhysRevA.74.032307,PhysRevLett.121.070402}, topological phenomena~\cite{PhysRevA.82.033429,PhysRevA.96.033846}, nonlinear dynamics~\cite{PhysRevA.101.023802,PhysRevA.106.062407}, interacting systems~\cite{PhysRevA.101.023802,PhysRevA.106.062407}, decoherence properties~\cite{PhysRevA.67.042315,PhysRevLett.106.180403}, and complex disordered systems~\cite{PhysRevLett.111.180503,PhysRevB.96.144204,PhysRevLett.123.140501,PhysRevResearch.3.023052}.

In parallel with the various theoretical proposals, DTQWs have been experimentally demonstrated in a number of platforms, e.g., trapped ions~\cite{PhysRevLett.103.090504}, trapped atoms~\cite{karski2009quantum}, photons in waveguides~\cite{PhysRevLett.100.170506}, light using optical devices~\cite{PhysRevResearch.4.023042}, and even in nuclear magnetic resonance systems~\cite{PhysRevA.67.042316,PhysRevA.72.062317} and superconducting qubits~\cite{PhysRevX.7.031023}.
These experimental demonstrations highlight the versatility and wide-ranging applications of DTQWs.	
On the other hand, a major difficulty in practical implementations is to preserve quantum coherence for long periods of time, due to the high sensitivity of the quantum states and decoherence process induced by interaction with the environment~\cite{n3,PhysRevA.76.032111}.
%
%
%

Recently, disordered DTQWs have garnered significant attention due to their rich dynamical properties~\cite{ahlbrecht2012asymptotic,PhysRevLett.111.180503,PhysRevA.89.042307,PhysRevE.100.032106,wang2018dynamic,PhysRevA.96.033846,PhysRevB.96.144204,PhysRevLett.123.140501,wang2018dynamic,ahlbrecht2012asymptotic,zeng2017discrete,n5}. 
We remark, for example, that disorder in DTQWs can lead to Anderson localization, characterized by exponentially localized eigenstates of a quantum particle~\cite{n1,n2}.
%
Conversely, studies have also shown that DTQWs with temporal inhomogeneity may exhibit diffusive dynamics similar to classical random walks~\cite{PhysRevLett.106.180403}. 
%
%

In addition, Ahlbrecht and coauthors have investigated~\cite{ahlbrecht2012asymptotic} the influence of random quantum coins with spatial and temporal dependence that act simultaneously on the dynamics of a DTQW, and reported the presence of diffusive dynamics in the long-time limit.
More recently, Mendes and coauthors~\cite{PhysRevE.99.022117} have studied numerically the effects of static long-range correlations applied to the conditional displacement operator in a Hadamard quantum walk, characterizing a localized-delocalized state transition with onset of ballistic dispersion controlled by the  parameter that adjusts the degree of correlation.

In DTQWs the insertion of random inhomogeneities in the time evolution operator not only affects the transport properties, but also changes the coin-position entanglement features~\cite{PhysRevLett.111.180503,PhysRevA.89.042307,PhysRevE.100.032106,wang2018dynamic}.
In~\cite{PhysRevA.89.042307} the effects of different types of disorder on the generation of coin-position entanglement have been investigated, with the static disorder shown not to be a very efficient mechanism to generate quantum entanglement. 
In contrast, for any initial condition, dynamic and fluctuating disorder leads to maximum entangled states asymptotically in time. 
Recently, we have shown~\cite{PhysRevE.100.032106} that static inhomogeneities with aperiodic correlations can also give rise to maximally entangled~states. 

In general, understanding the time evolution of DTQWs in the presence of disorder sources such as noise, fluctuations, and random inhomogeneity is crucial for practical implementations, as it enables greater control over the dynamics in the presence of environmental interactions.

In this work, we investigate the role of random inhomogeneities with long-range spatial and temporal correlations on the quantum walk dynamics.
We study the time evolution of an initial qubit state following the DTQW protocol, with long-range (power-law) correlations displaying space-time dependence (static and dynamic inhomogeneities) encoded in the phases of the quantum coin.
Interestingly, depending on the relative strength of the correlation parameters, quite diverse dynamics arise in the quantum system, ranging from localized to superdiffusive (including ballistic) behavior.
Overall, our findings advance in understanding the interplay between the effect of inhomogeneous correlations and the resulting dynamics of DTQWs, possibly bringing about not only theoretical gain but eventually practical relevance as well.

The article is organized as follows.
In Section~II we introduce the model and describe the general formalism. 
Results and discussion are presented in Section~III.
Lastly, final remarks and conclusions are left to Section~IV.

\section{Model and formalism}

We consider a quantum random walker propagating in a one-dimensional (1D)~lattice with $N$ sites, discrete positions indexed by integers $n \; (= 1, 2, ..., N)$, and long-range spatial-temporal correlated inhomogeneities introduced as described below. 
The walker is a qubit with internal degree of freedom spanned by a two-level system that defines the basis of the so-called coin space~\cite{n4}: 
$\mathcal{H}^\mathcal{C}\equiv\{\ket{\uparrow}=(1,0)^{T}$, $\ket{\uparrow}=(0,1)^{T}\}$, where $T$~denotes transpose. 
The qubit state $\ket{\Psi}$ belongs to a Hilbert space set by the tensor product of two spaces, $\mathcal{H}=\mathcal{H}^\mathcal{P}\otimes \mathcal{H}^\mathcal{C}$, with $\mathcal{H}^\mathcal{P}$ assigned to the position space consisting of states~$\{|n\rangle\}$. 
The generic initial $(t = 0)$ state of the quantum walker is written as the superposition
\begin{eqnarray}
\label{equation1}	
\ket{\Psi_0} \equiv \ket{\Psi(t=0)} = \sum_{n}\big ( a_{n,t=0}\ket{\uparrow} +b_{n,t=0}\ket{\downarrow}\big )\otimes|n\rangle,
\end{eqnarray}
with normalization $\sum_{n}\big (|a_{n,t=0}|^{2}+|b_{n,t=0}|^{2} \big )=1$. 

The system evolution with discrete time~$t$ depends on both internal and spatial degrees of freedom, which are respectively driven by the unitary operators $\hat{C}$ (quantum coin) and $\hat{S}$ (conditional displacement operator).
In a general description, one can express~\cite{n4} a single-site quantum coin as an arbitrary unitary SU(2) matrix on the basis of the coin space, 
\begin{eqnarray}		
\label{eq_quantumcoin}
\hat{C}(q,\theta,\phi)= \left(\begin{array}{cc} \sqrt{q} & \sqrt{1-q}e^{i\theta}\\ 
\sqrt{1-q}e^{i\phi} & -\sqrt{q}e^{i(\theta+\phi)}
\end{array} \right), 
\end{eqnarray}
where the angles $0\leq\theta \leq 2\pi$ and $0 \leq \phi\leq 2\pi$ control the relative phase between the two coin states, while the parameter $q\in[0,1]$ drives the spatial bias of the quantum coin. 
For example, for $q=1/2$ and $\theta=\phi=0$ one~has a fair quantum coin that chooses both possible directions in the 1D~lattice (left or right) with equal probability (Hadamard coin)~\cite{n4}. 
In this work, we set $q=1/2$ and consider the stochastic evolution of the random phases~$\theta$ and~$\phi$ as defined below.

On the other hand, the conditional displacement operator, 
\begin{eqnarray}
\label{shiftoperator}
\hat{S}= \sum_{n}\big (\ket{\uparrow}\bra{\uparrow}\otimes |n+1\rangle\langle n| + \ket{\downarrow}\bra{\downarrow} \otimes |n-1\rangle\langle n| \big ), 
\end{eqnarray}
does not alter the walker's internal state, but moves it from position $n$ to $n+1$ $(n-1)$ if the internal state is $\ket{\uparrow}$ $(\ket{\downarrow})$. 
The system evolution with discrete time~$t$ from the initial state $\ket{\Psi_0}$, Eq.~(\ref{equation1}), is thus obtained through $\ket{\Psi(t)}=(\hat{U})^t \ket{\Psi_0}$, so that the time evolution operator $\hat{U}=\hat{S}\hat{C}$ describes the simultaneous action on the quantum walker of both quantum coin and conditional displacement operators.

In order to introduce random inhomogeneity effects and spatial and temporal long-range correlations in the  DTQW model, 
%
%
we first describe a general procedure to generate a set $\{ \tilde{V}_{n} \}$ of random variables with heterogeneous distributions and long-range correlations.
We start by considering a large number $M \gg 1$ of generic random variables~$\tilde{V}_{j}$ given by the sum, 
\begin{eqnarray}
\label{eq_correlation}
\tilde{V}_{j} =\sum_{k=1}^{M/2} \left[\left(\frac{2\pi}{M}\right)^{(1-\nu)}\frac{1}{k^{\nu}}\right]^{1/2} \cos\left(\frac{2\pi jk}{M} + \mu_{k} \right),
\end{eqnarray}
where $j = 1, 2, ..., M$, with $\mu_{k}$ denoting $M/2$ independent random phases uniformly distributed in the interval $[0,2\pi)$.
We note that $\tilde{V}_{j}$ corresponds~\cite{mouralyra,PhysRevE.99.022117} to the trace of a fractional Brownian motion, with the sequence of values $\{ \tilde{V}_{j} \}$ displaying asymptotic power-law spectrum in the form $S(k) \sim 1/k^\nu$. 
The parameter $\nu \ge 0$ controls the degree of correlation of the set of variables $\tilde{V}_{j}$.
Indeed, for $\nu = 0$ the values $\{ \tilde{V}_{j} \}$ are essentially uncorrelated, whereas for $\nu > 0$ they present long-range correlations even for large~$M$. 
Also, Eq.~(\ref{eq_correlation}) yields a $j$-dependent probability distribution for each variable $\tilde{V}_{j}$, therefore leading to a statistically inhomogeneous (or heterogeneous) set of random variables $\tilde{V}_{j}$. 

From the averages $\overline{\cos\mu_{k}}=\overline{\sin\mu_{k}}=0$, we notice that the first and second moments read $\overline{\tilde{V}}_{j}=0$ and $\overline{\tilde{V}_j^{2}}=1$, independent of~$M$ and~$\nu$. 
By next considering
\begin{equation}
%
%
V_{j}= \pi \left [ \tanh(\tilde{V}_{j})+1 \right ], 
\label{new2}
\end{equation}
we constrain the normalized variables $V_j$ to lie within the range $[0,2\pi)$.
We remark that transformation~(\ref{new2}) does not change the asymptotic power-law behavior of the correlations of the sequence~\cite{PhysRevE.99.022117}.

\begin{figure}
\centering
\includegraphics[width=\linewidth]{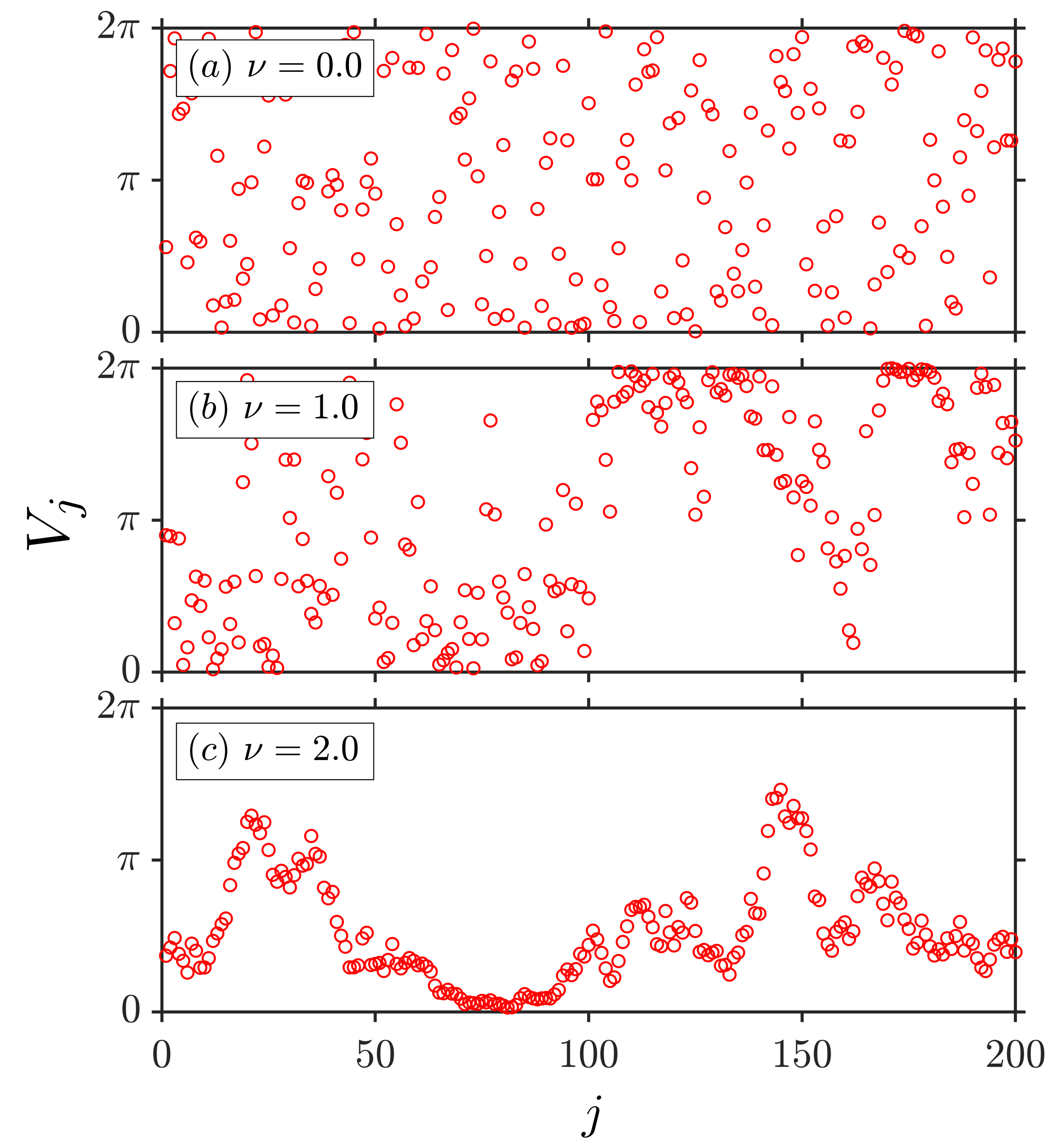}
\caption{Single realizations of the distribution of values of the generic random variables $V_{j}$ as a function of $j \; (= 1, 2, ..., M)$, with $M = 200$, generated from Eqs.~(\ref{eq_correlation}) and~(\ref{new2}) for three representative values of the power-law exponent: (a) $\nu=0$, (b)~$\nu = 1.0$, and (c) $\nu = 2.0$. 
When $\nu=0$, we obtain the uncorrelated case, with a uniformly random distribution of $V_{j}$-values. 
For increasing $\nu > 0$ the sequences $\{ V_j \}$ smooth out towards the profile of a fractional Brownian motion with power-law spectrum $S(k) \sim 1/k^\nu$.
Variables $V_j$ are used in this work to generate distributions of long-range correlated inhomogeneous coin phases $\theta_t$ and~$\phi_n$, by taking $j \to t$ and~${j \to n}$, respectively.}
%
%
\label{figure1}
\end{figure}

Figure~\ref{figure1} shows profiles of single realizations of the generic variables $V_j$ generated from Eqs.~(\ref{eq_correlation}) and~(\ref{new2}), for $M=200$ and three representative values of the power-law exponent~$\nu$. 
In Fig.~\ref{figure1}(a) we set $\nu=0$ and obtain the uncorrelated case, with uniformly distributed random $V_{j}$-values. 
%
%
On the other hand, for increasingly positive values of~$\nu$, Figs.~\ref{figure1}(b)-(c) indicate that the patterns of $V_{j}$-values gradually smooth out, resembling the profile of the trace of a fractional Brownian motion with power-law spectrum $S(k) \sim 1/k^\nu$~\cite{mouralyra}. 
%
%
%

%
%
%
%
%
%
%

At this point, the connection with the DTQW model can be built. 
We aim to introduce temporal and spatial inhomogeneities in the quantum coin operator $\hat{C}$, Eq.~(\ref{eq_quantumcoin}).
By fixing ${q = 1/2}$, two degrees of freedom are left, associated with the phases~$\theta$ and~$\phi$.
One possible choice to yield time and space dependence in the coin operator is to set $\theta \to \theta_t$ and $\phi \to \phi_n$, respectively, so that $\hat{C}(q,\theta,\phi) \to \hat{C}(1/2,\theta_t,\phi_n)$ in Eq.~(\ref{eq_quantumcoin}).
Now, Eqs.~(\ref{eq_correlation}) and~(\ref{new2}) can be used to generate long-range correlated random sequences of both coin phases. 
For example, by assigning the general index $j$ to the discrete time~$t$, we set $V_j \to \theta_t$ along with $M \to T$ (maximum time considered) and the power-law exponent $\nu \rightarrow \alpha_{t}$.
Conversely, by associating $j$ with the lattice site~$n$, we take $V_j \to \phi_n$ along with $M \to N$ and $\nu \rightarrow \beta_{s}$. 
It is also important to mention that each sequence $\{\theta_t,\phi_n\}$, for fixed $\alpha_t \ge 0$ and $\beta_t \ge 0$, is generated using statistically independent sets $\{ \mu_{k} \}$.
%
%
This procedure ultimately leads to long-range correlations with temporal inhomogeneities in the $\theta$-phase (driven by $\alpha_t$) and spatial inhomogeneities in the $\phi$-phase (driven by $\beta_s$).

The time evolution protocol is described as follows.
The state of the qubit after $t$ discrete time steps can be expressed as 
%
\begin{eqnarray}
\label{eq6}
|\Psi (t) \rangle = (\hat{U})^t \ket{\Psi_0} = \sum_{n} \left(\psi_{t,n}^{\uparrow}\ket{\uparrow}+\psi_{t,n}^{\downarrow}\ket{\downarrow}\right),
\end{eqnarray} 
where $\psi_{t,n}^{\uparrow}$ and $\psi_{t,n}^{\downarrow}$ are the probability amplitudes of obtaining the internal states $\ket{\uparrow}$ and $\ket{\downarrow}$ at the position~$n$ in time~$t$.
With the use of Eqs.~(\ref{equation1})-(\ref{shiftoperator}), together with $\hat{C}(q,\theta,\phi) \to \hat{C}(1/2,\theta_t,\phi_n)$, these amplitudes are given in terms of their values in the preceding time and neighbor sites, along with the coin phases $\{ \theta_t, \phi_n \}$, by the following recurrence relations, 
\begin{eqnarray}
\label{eq7}
& &\psi_{t,n}^{\uparrow} = \frac{1}{\sqrt{2}}\left(\psi_{t-1,n+1}^{\uparrow}+e^{i\theta_t}\psi_{t-1,n+1}^{\downarrow}\right),\nonumber \\ 
& &\psi_{t,n}^{\downarrow}=\frac{1}{\sqrt{2}}\left(e^{i\phi_n}\psi_{t-1,n-1}^{\uparrow} -e^{i(\theta_t+\phi_n)} \psi_{t-1,n-1}^{\downarrow}\right).
\end{eqnarray}
These equations are then iterated numerically, giving rise to the quantum state~(\ref{eq6}) in the subsequent time, and~so~on.

\begin{figure*}[t]
\centering		\includegraphics[width=\linewidth]{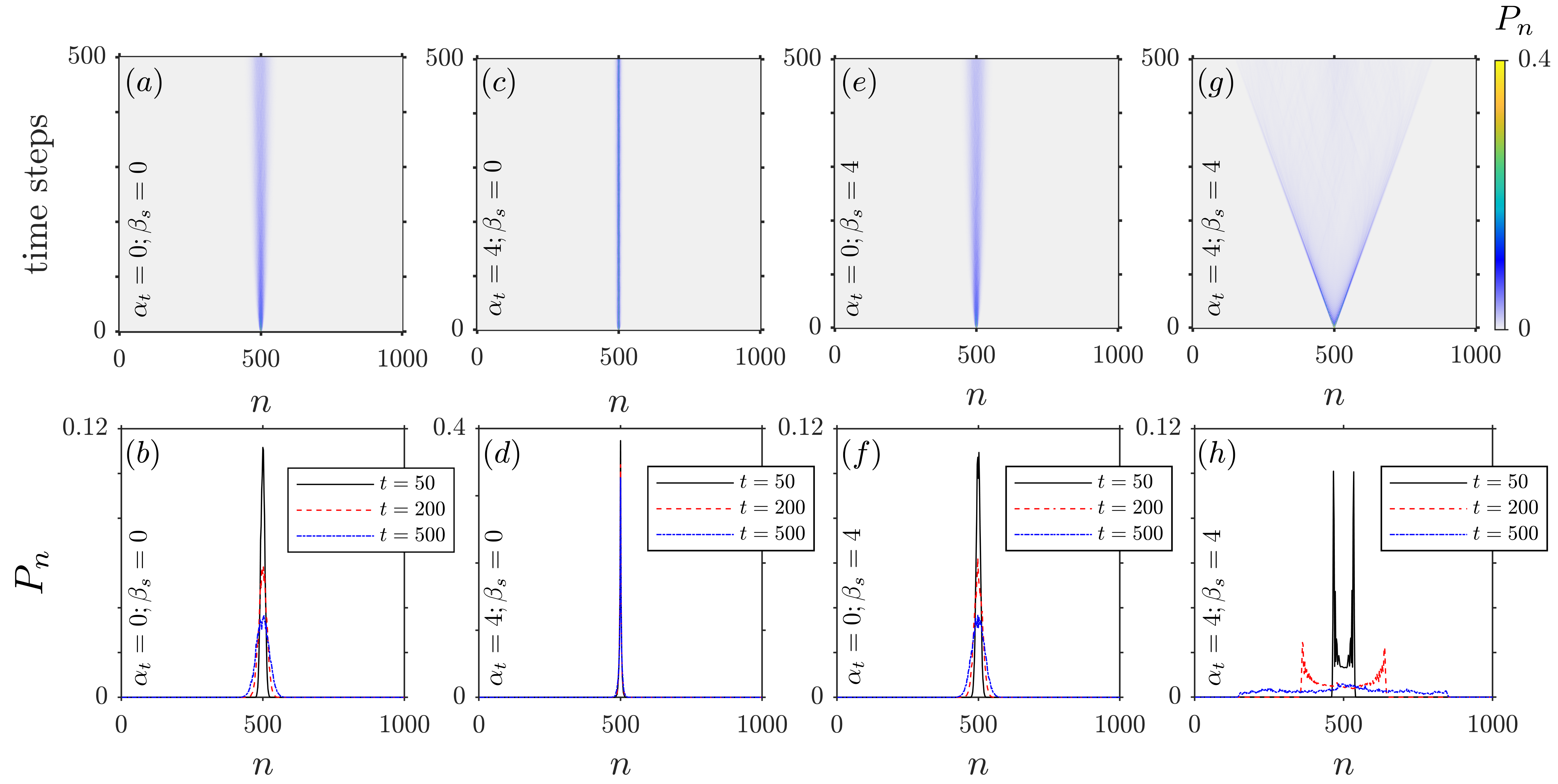}
\caption{Dynamic evolution of the probability $P_n(t)$ of finding the quantum walker at the site~$n$ in time~$t$, in a 1D lattice with $N = 1000$ sites and for a maximum time $T = N/2$. 
The upper panel shows $P_n(t)$ in the space-time plane, whereas the lower panel displays snapshots in $t=50, 200, 500$.
We consider four relevant choices of the parameters $\{ \alpha_{t}, \beta_{s}\}$ that drive the temporal and spatial degrees of correlation and inhomogeneity in the phases $\{\theta,\phi\}$ of the quantum coin:
(a)-(b) $\alpha_{t}=\beta_{s}=0$ (uncorrelated inhomogeneties in both space and time); 
(c)-(d) $\alpha_{t}=4$ and $\beta_{s}=0$ (strong long-range temporal correlation and uncorrelated spatial inhomogeneity);
(e)-(f) $\alpha_{t}=0$ and $\beta_{s}=4$ (uncorrelated temporal inhomogeneity and strong spatial correlation);
(g)-(h) $\alpha_{t}=\beta_{s}=4$ (strong long-range correlated inhomogeneties in both space and time). 
%
%
%
Different dynamical scenarios emerge by adjusting the combination of these  parameters (see text).}
%
%
\label{figure2}
\end{figure*}

The dynamical analysis of the DTQW can be performed through the study of the propagation of the qubit wave packet. 
From the probability of finding the quantum walker at the site $n$ in time $t$, $P_n(t) = {|\bra{\Psi (t)}\uparrow,n\rangle|^{2} +} |\bra{ \Psi(t)}\downarrow,n\rangle|^{2}$, we obtain its mean position as a function of time, $\overline{n}(t)=\sum_{n}nP_{n}(t)$, and the associated dispersion,  
\begin{eqnarray}
\label{eq_meandisplacement}
\sigma(t)=\sqrt{\sum_{n} \big [n-\overline{n}(t) \big ] ^{2}P_{n}(t)}.
\end{eqnarray}
In general terms, the asymptotic relation $\sigma (t) \sim t^H$ between the dispersion and Hurst exponent~$H$ quantifies key aspects of the walker's dynamics~\cite{nn1}. 
For instance, normal diffusion of either classical (Brownian) or quantum random walkers is characterized by the Hurst exponent ${H = 1/2}$, with statistics governed by the central limit theorem (CLT). 
On the other hand, anomalous diffusion presents ${H \not = 1/2}$, for example, subdiffusion ${(0 < H < 1/2)}$ and superdiffusion $(H > 1/2)$ processes, including the ballistic $(H = 1)$ and even superballistic $(H > 1)$ cases.
In particular, the interest in anomalous superdiffusive processes has grown considerably in the last decades~\cite{nn1}, as superdiffusivity has been increasingly reported in many distinct contexts, usually related to the generalized CLT, L\'evy distributions, and extreme event statistics, from anomalous quantum transport and quantum work~\cite{nn2,nn2b,nn2c}, to photons propagating in random lasers~\cite{nn3,nn4,nn5,nn6}, and efficient random searches~\cite{nn7,nn8,nn9}, to name a few. 

Our resuts presented in the next section characterize diverse dynamical regimes displayed by long-range correlated inhomogeneous DTQWs. 

\section{Results and discussion}

Results were obtained following the numerical time evolution protocol (Eqs.~(\ref{eq6}) and~(\ref{eq7})) applied to a quantum walker initially located at the nearly symmetric position ${n_{0}=N/2}$, with the initial state $\ket{\Psi_{0}}$ displaying equiprobable internal states $\ket{\uparrow}$ and  $\ket{\downarrow}$, 
%
%
\begin{eqnarray}
\label{initial_state}
\ket{\Psi_{0}}=\frac{1}{\sqrt{2}}(\ket{\uparrow}+i\ket{\downarrow})\otimes \ket{n_{0}}.
\end{eqnarray}
This means that the initial probability $P_n(t=0)$ of finding the walker at sites $n \not = N/2$ is null.
For $t > 0$, the recursive application of the conditional displacement operator, Eq.~(\ref{shiftoperator}), can yield a non-null $P_n(t)$ for $n \not = N/2$, as also seen from the probability amplitudes, Eq.~(\ref{eq7}).  
However, due to the stochastic character of the sequence $\{ \theta_t, \phi_n \}$, the walker's dynamics is strongly dependent on the choice of parameters $\{ \alpha_t, \beta_s \}$ that drive the degree of temporal and spatial correlations. 
Accordingly, the range of sites around the starting point with considerable probability of finding the walker in a given time is also greatly influenced by the dispersion properties (e.g., diffusive, superdiffusive, ballistic, etc.; see below).  
In most cases, we consider a maximum time $t = T$ not too large, so that the boundaries of the 1D~lattice at $n = 1$ and $n = N$ are not accessed, thus avoiding edge effects.  
Also, for each choice $\{ \alpha_t, \beta_s \}$ averages were taken over $5000$~independent random realizations of the sequence~${\{ \theta_t, \phi_n \}}$.

We start with the time evolution of the probability~$P_{n}(t)$ in a 1D~lattice with $N=1000$ sites, for a maximum time $T = N/2$. 
Figure~\ref{figure2} shows $P_{n}(t)$ for four relevant choices of $\{\alpha_{t},\beta{s}\}$.  
In each case, the upper panel displays the evolution profile of the probability in the space-time plane, while the lower panel presents snapshots of $P_{n}(t)$ in $t=50, 200, 500$. 
For uncorrelated inhomogeneities ($\alpha_{t}=\beta_{s}=0$, Figs.~\ref{figure2}(a)-\ref{figure2}(b)), the coin phases $\{\theta,\phi\}$ are randomly distributed in time and space, respectively. 
We observe in Fig.~\ref{figure2}(a) that the dynamics of the wave packet profile indicates a low degree of mobility of the quantum walker, which spreads slowly over the 1D~lattice. 
The probability snapshot shows that the qubit wave packet acquires a Gaussian profile (blue line curve in Fig.~\ref{figure2}(b)) that resembles the propagation of a classical (Brownian) random walker (see, also, the dispersion results below). 
%
%

On the other hand, for $\alpha_{t}=4$ (strong long-range temporal correlation)  and $\beta_{s}=0$ (uncorrelated spatial inhomogeneity), Figs.~\ref{figure2}(c)-\ref{figure2}(d) show that the probability profile of the qubit wave packet remains trapped around the initial position $n_{0}=N/2=500$, in a picture consistent with a localized quantum state (see below).
%
%
In contrast, by setting $\alpha_{t}=0$ (uncorrelated temporal inhomogeneity) and $\beta_{s}=4$ (strong spatial correlation) in Figs.~\ref{figure2}(e)-\ref{figure2}(f), a probability function with Gaussian profile is recovered, including a spread pattern in Fig.~\ref{figure2}(e) similar to that observed in the fully uncorrelated case, Fig.~\ref{figure2}(a). 
These findings suggest that the time-inhomogeneity aspects in the coin phases seem to prevail over the spatial fluctuations in what concerns the quantum walker's dynamics. 

At last, when both phases are tunned in the strong correlation regime, $\alpha_{t} = \beta_{s} = 4$, the above scenarios change drastically, as seen in Figs.~\ref{figure2}(g)-\ref{figure2}(h).  
The qubit wave packet spreads rather fast, with much stronger dispersive character. 
The probability profile exhibits two nearly symmetric peaks that decay monotonically with time, see Fig.~\ref{figure2}(h), thus suggesting a delocalization of the quantum walker, with two most probable positions equidistant from the starting point. 

\begin{figure}[!t]
\centering
\includegraphics[width=\linewidth]{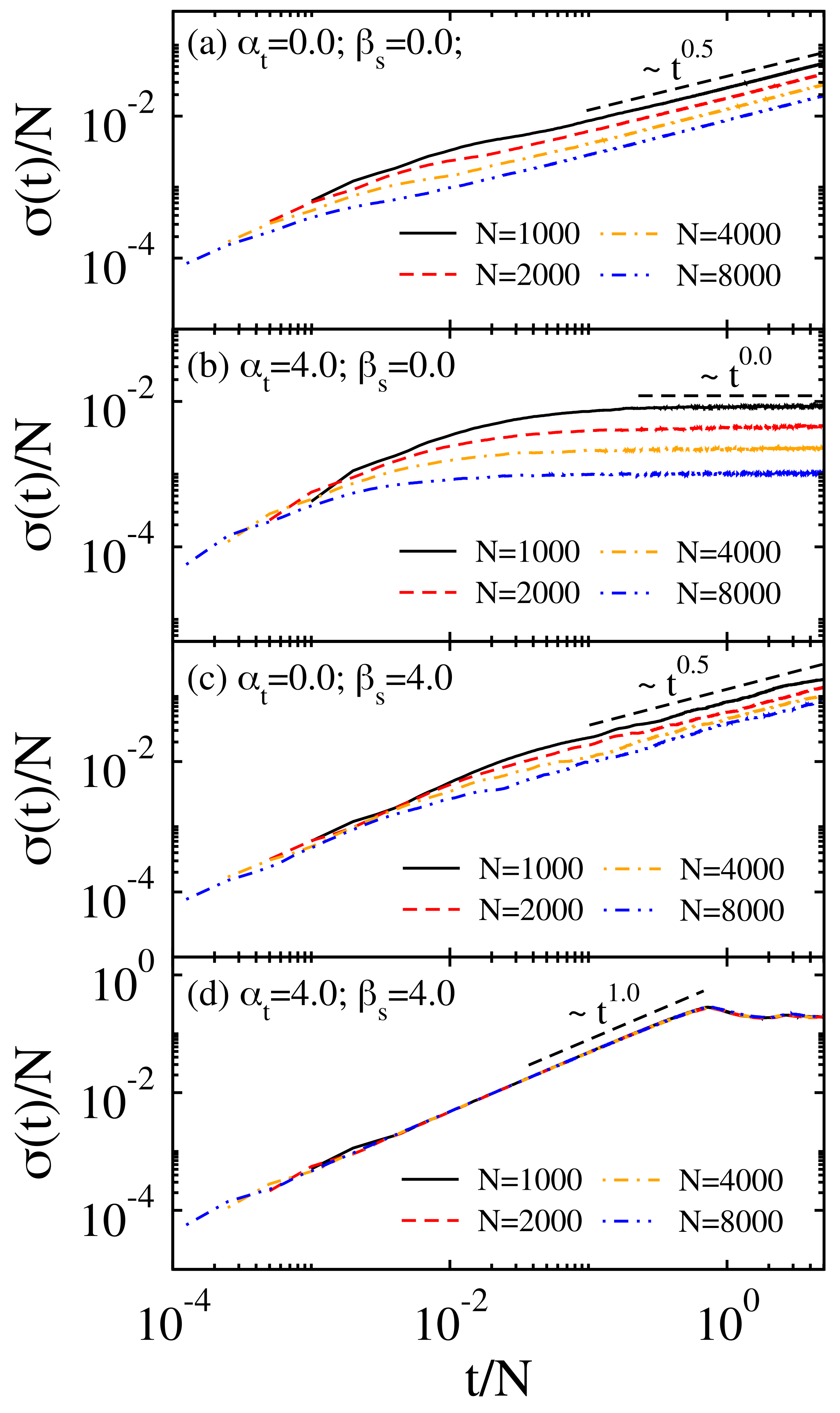}
\caption{Normalized dispersion, $\sigma (t) /N$, as a function of the normalized discrete time, $t/N$, for the same parameter choices $\{ \alpha_t, \beta_s \}$ of Fig.~\ref{figure2}, lattice sizes $N = 1000, 2000, 4000, 8000$, and maximum time $T = 5N$.  
The asymptotic behavior $\sigma (t) \sim t^H$, where $H$ is the Hurst exponent, defines the localized ${(H \sim 0)}$, diffusive $(H = 1/2)$, and superdiffusive (ballistic) $(H = 1)$ dynamics of the quantum walker, consistently with the $P_n(t)$ results of~Fig.~\ref{figure2}.} 
%
%
%
%
\label{figure3}
\end{figure}

Figure~\ref{figure3} presents results of the normalized dispersion, $\sigma (t) / N$, given in Eq.~(\ref{eq_meandisplacement}), for the same parameter choices of Fig.~\ref{figure2}. 
In addition to the lattice with $N = 1000$ sites, we also show data for $N = 2000, 4000, 8000$, with a larger maximum time, $T = 5N$. 
We notice that the results in Fig.~\ref{figure3} are fully consistent with those of Fig.~\ref{figure2}. 
For example, after an initial  transient, the uncorrelated case ($\alpha_{t}=\beta_{s}=0$), shown in Fig.~\ref{figure3}(a), displays for all~$N$ an asymptotic  scaling relation $\sigma (t) \sim t^{0.5}$, which is consistent with the Brownian-like Hurst exponent $H = 1/2$, typical of the normal dynamics of a classical random walker driven by a Gaussian probability function and~CLT. 

A similar picture with $H = 1/2$ is also observed in Fig.~\ref{figure3}(c) for the case with uncorrelated temporal inhomogeneity and strong long-range spatial correlation, $\alpha_t = 0$ and $\beta_s = 4$, in agreement with Figs.~\ref{figure2}(e)-\ref{figure2}(f). 
These results indicate that in the absence of temporal correlations, $\alpha_{t}=0$, the qubit wave packet exhibits diffusive behavior, regardless of the value of $\beta_{s}$.

Interestingly, the localized scenario of Figs.~\ref{figure2}(c)-\ref{figure2}(d), with strong temporal correlation and uncorrelated spatial inhomogeneity, $\alpha_t = 4$ and $\beta_s = 0$, reveals an asymptotic saturation of the dispersion, $\sigma \sim t^0$, for long times and all~$N$. 
The saturation value of $\sigma$ depends on the size~$N$ of the 1D~lattice. 
We thus notice that the disorder effects in the form of the joint action of strong long-range temporal correlation and lack of spatial correlation in the quantum coin cause the qubit states to display an Anderson-like localization behavior.
In this case, the random scattering of the qubit wave packet, driven by the stochastic sequence $\{ \theta_t, \phi_n \}$, promotes a significant spatial confining of the quantum walker around the initial position. 
%

%

%
%

%
%

\begin{figure}[!t]
\centering
\includegraphics[width=\linewidth]{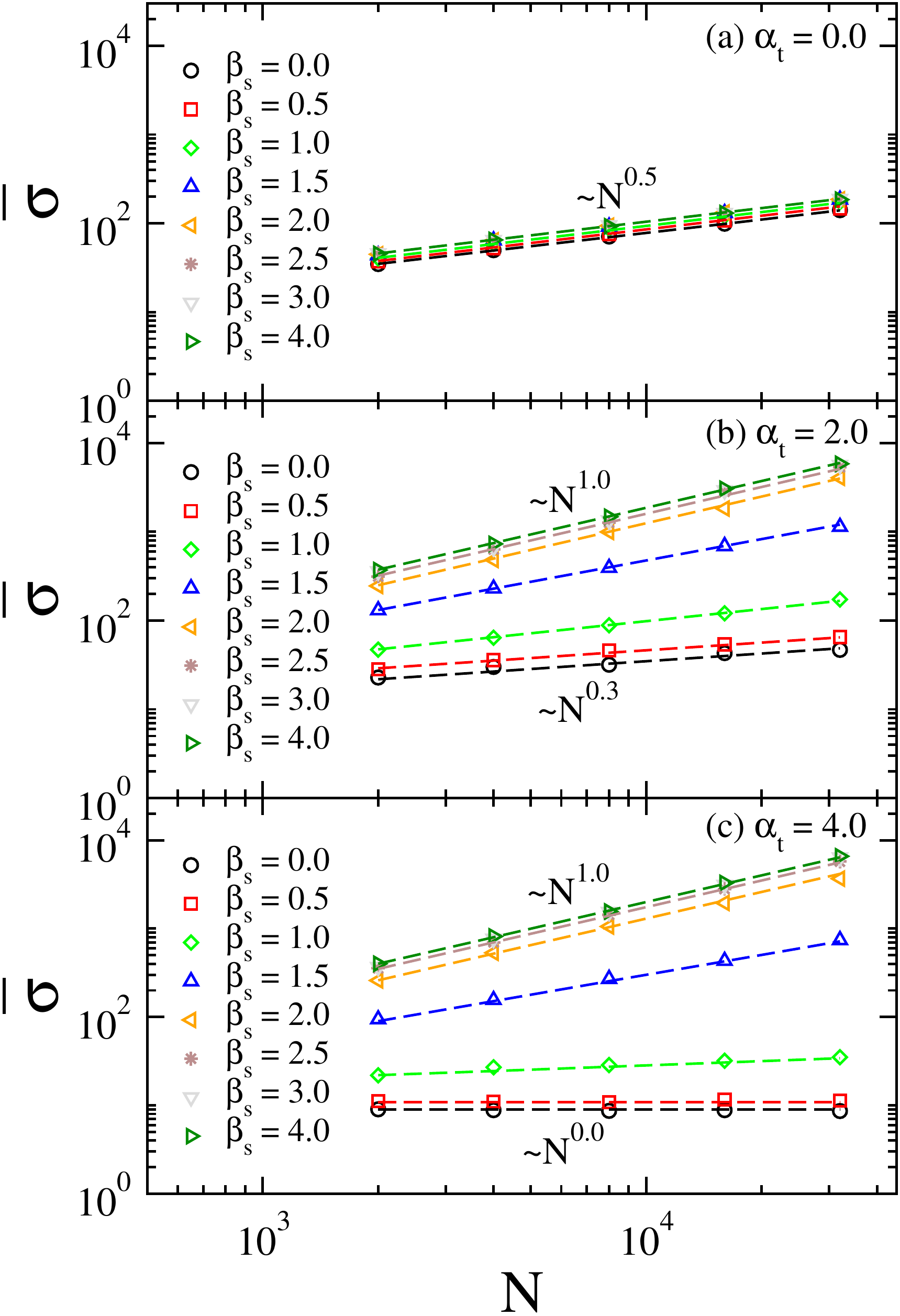}
\caption{Long-time average dispersion $\overline{\sigma}$ as a function of the system size~$N$. 
Parameter $\alpha_{t}$ that controls the temporal correlations is fixed at the values (a)~$\alpha_{t} = 0$, (b)~$\alpha_{t} = 2.0$, and (c)~$\alpha_{t} = 4.0$, while parameter~$\beta_{s}$ driving the spatial correlations varies in the range $[0,4.0]$.
The scaling behavior $\overline{\sigma} \sim N^\gamma$ characterizes a variety of dynamical regimes, 
in agreement with Figs.~\ref{figure2}-\ref{figure3}.
%
%
%
(a) In the uncorrelated temporal case, $\alpha_{t}=0$, the dynamics is independent of the spatial correlation degree, exhibiting diffusive behavior ($\gamma=1/2$) for all~$\beta_{s}$. 
(b)~For an intermediate degree of temporal correlation, $\alpha_{t}=2.0$, the dynamics ranges from subdiffusive $(0 < \gamma < 1/2)$ to superdiffusive $(\gamma > 1/2)$, and up to ballistic $(\gamma =1.0)$. 
(c) For strong temporal correlations, $\alpha_{t}=4.0$, all regimes can be accessed, from localized $(\gamma \sim 0)$ to ballistic $(\gamma =1.0)$, as $\beta_{s}$ increases from $\beta_{s} = 0$ to~${\beta_{s} = 4.0}$.}
\label{figure4}
\end{figure}

\begin{figure}[!t]
\centering		\includegraphics[width=\linewidth]{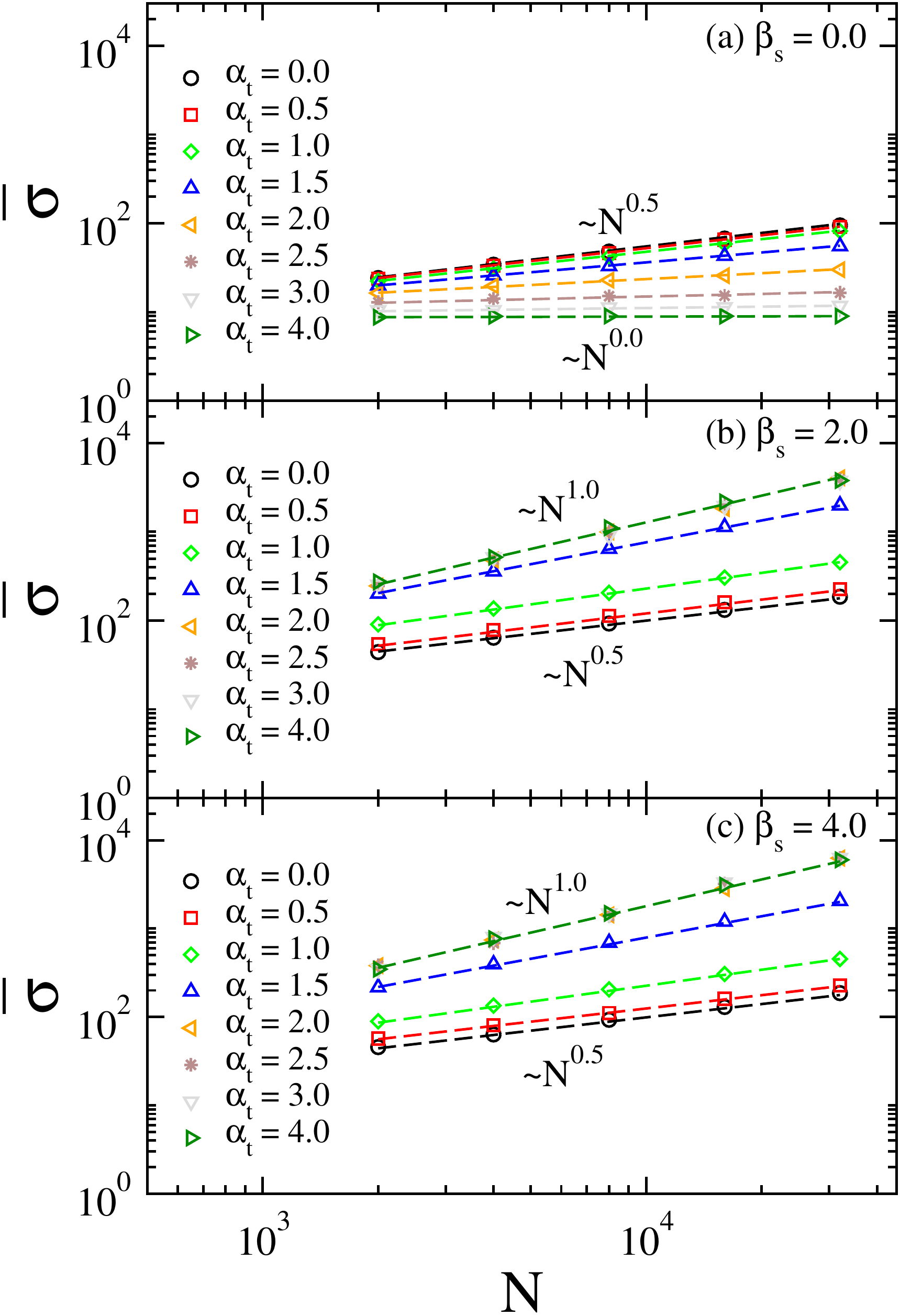}
\caption{Long-time average dispersion $\overline{\sigma}$ as a function of the system size~$N$, depicting the counterpart of Fig.~\ref{figure4}, but with $\beta_{s}$ fixed at the values (a)~$\beta_{s} = 0$, (b)~$\beta_{s} = 2.0$, and (c)~$\beta_{s} = 4.0$, while $\alpha_{t}$ varies in the range $[0,4.0]$.
The scaling behavior  $\overline{\sigma} \sim N^\gamma$
characterizes diverse dynamic regimes identified by the $\gamma$-value, consistent with Figs.~\ref{figure2}-\ref{figure4}.}
%
%
\label{figure5}
\end{figure}
%

%
%
%

On the other hand, the fast spread of the wave packet shown Figs.~\ref{figure2}(g)-\ref{figure2}(h) for long-range spatial and temporal correlations, $\alpha_{t} = \beta_{s} = 4$, yields a superdiffusive (ballistic) dynamic behavior, $\sigma (t) \sim t^{1.0}$ for all~$N$, with the dispersion increasing linearly with time and Hurst exponent~$H=1$, up to times $t \sim N/2$ that mark the onset of the reaching of the lattice boundaries (this is, in fact, the only result in which the walker hits the borders).

In order to offer a complementary analysis, Fig.~\ref{figure4} and Fig.~\ref{figure5} display the dispersion $\overline{\sigma}$ averaged over the last~100~time steps (i.e., from $t = T - 100$ to $t = T$, with $T = N/2$), as a function of the lattice size~$N$, up to a much larger value, $N = 32000$.
In a way similar to the Hurst exponent, we define the relation 
$\overline{\sigma} \sim N^{\gamma}$, where the exponent~$\gamma$ quantifies the scaling with the system size~$N$ of the width of the qubit wave packet in the asymptotic long-time regime. 
We thus can see in more detail how the competition between long-range correlations in time and space can lead to a quite rich dynamics of the quantum walker, as progressively larger lattices are considered. 
%

%
%

%
Figure~\ref{figure4}(a) shows the average dispersion $\overline{\sigma}$ when the inhomogeneity is uncorrelated in time, $\alpha_{t}=0$, for various degrees of spatial correlation. 
In this case, the quantum walker exhibits diffusive dynamics ($\gamma = 1/2$) for all $\beta_s$-values, confirming the previous results of Fig.~\ref{figure2} and Fig.~\ref{figure3}.
On the other hand, intermediate long-range temporal correlations, $\alpha_{t}=2.0$ in Fig.~\ref{figure4}(b), induce a rich spectrum of dynamic regimes, from subdiffusive $(0 < \gamma < 1/2)$ to ballistic $(\gamma = 1)$ as $\beta_s$ increases.
These regimes are also present for strong long-range temporal correlations, $\alpha_t = 4.0$ in Fig.~\ref{figure4}(c), with the addition of a localized behavior $( \gamma \sim 0)$ in the spatially uncorrelated case, $\beta_s = 0$.
%
%
%

On the other hand, Fig.~\ref{figure5} is the counterpart of Fig.~\ref{figure4}, but with the parameter $\beta_{s}$ fixed at the values 
$\beta_{s} = 0, 2.0, 4.0$, whereas $\alpha_t$ varies in the range $\alpha_t \in [0, 4.0]$. 
Essentially, the same dynamic regimes of Fig.~\ref{figure4} are also seen in Fig.~\ref{figure5}, as the degree of temporal correlation is increased for fixed $\beta_{s}$.
%
%
%
%
%
%
%

\begin{figure}[!t]
\centering
\includegraphics[width=\linewidth]{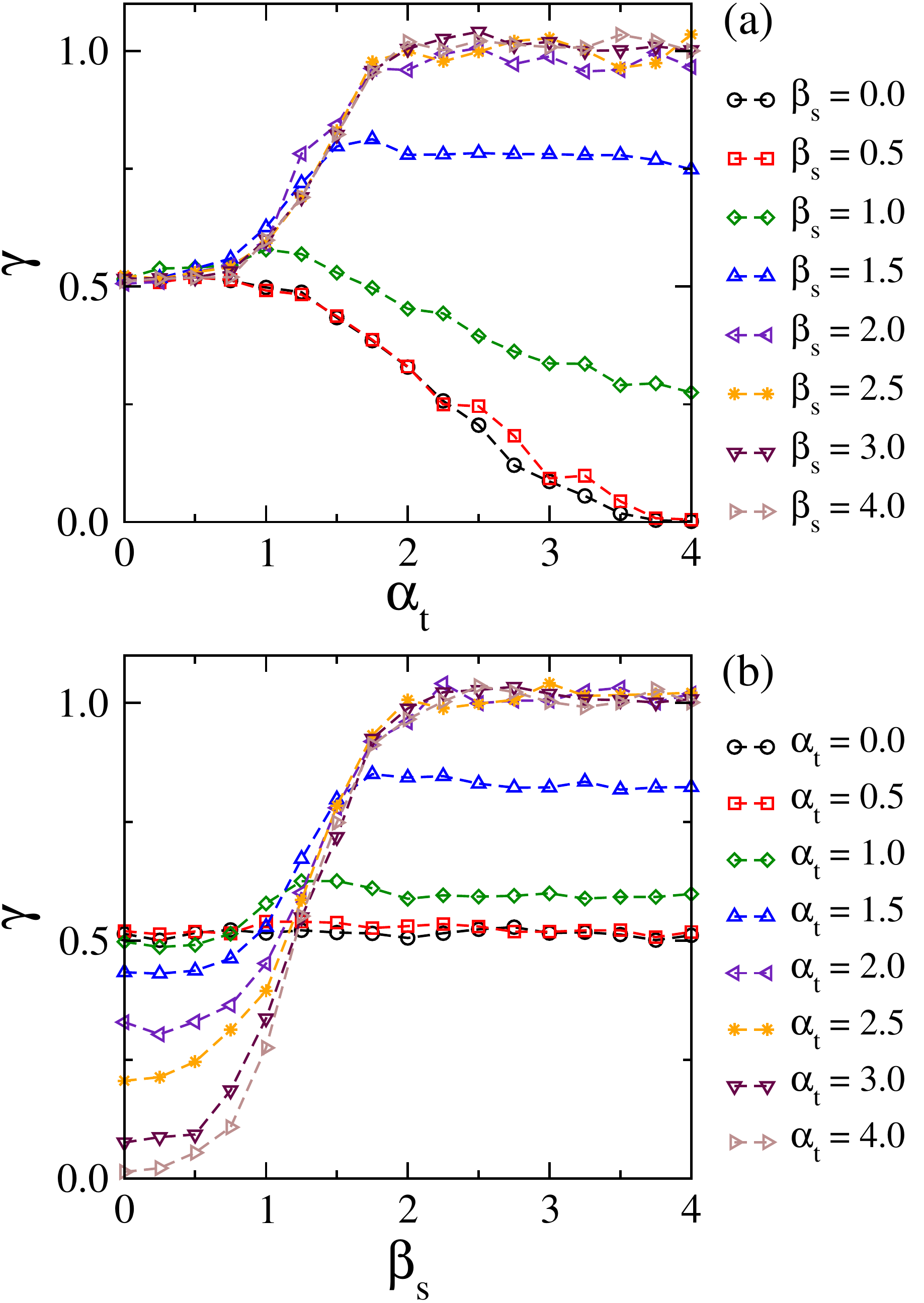}
\caption{Exponent $\gamma$ obtained from the scaling relation ${\overline{\sigma} \sim N^{\gamma}}$ of the average dispersion with the system size, for $N=2000, 4000, 8000, 16000, 32000$. 
The degree of temporal and spatial long-range correlations, set by $\alpha_t$ and $\beta_s$, respectively, drives the dynamics of the quantum walker in the localized $(\gamma \sim 0)$, subdiffusive $(0 < \gamma < 1/2)$, diffusive $(\gamma = 1/2)$, superdiffusive $(\gamma > 1/2)$, and ballistic $(\gamma = 1)$ regimes (see~text). 
%
%
%
%
%
%
%
%
%
%
}
\label{figure6}
\end{figure}

The values of the scale exponent $\gamma$ as a function of $\{\alpha_{t},\beta_{s}\}$ are plotted in Fig.~\ref{figure6}. 
Apart from small fluctuations, we note a roughly monotonic trend of $\gamma$ to either increase, decrease, or remain constant, when $\alpha_t$ or $\beta_s$ are fixed.
Instances of diffusive behavior $(\gamma = 1/2)$ are found for small $\alpha_t$ irrespective of the $\beta_s$-value (see, e.g., the nearly horizontal lines for $\alpha_t \lesssim 0.5$ in Fig.~\ref{figure6}(b)). 
On the other hand, for~${\alpha_{t} \gtrsim 1}$ the dynamic regime is strongly dependent on the spatial correlation degree, as seen in Fig.~\ref{figure6}(a), displaying a crossover from subdiffusive behavior for ${\beta_{s} \lesssim 1.0}$ to superdiffusive dynamics when ${\beta_{s} \gtrsim 1.0}$. 
In particular, ballistic behavior $(\gamma = 1)$ sets in only when both sources of inhomogeneities exhibit strong enough long-range correlations, i.e., for $\alpha_t \gtrsim 2.0$ and $\beta_s \gtrsim 2.0$.
Lastly, the localized regime $(\gamma \sim 0)$ of the quantum walker's wave function occupies a smaller fraction of the parameter space, being restricted to the range $\beta_s \lesssim 0.5$ with strong temporal correlations ($\alpha_t = 4.0$ in~Fig.~\ref{figure6}(b)).

\begin{figure}[!t]
\centering		\includegraphics[width=\linewidth]{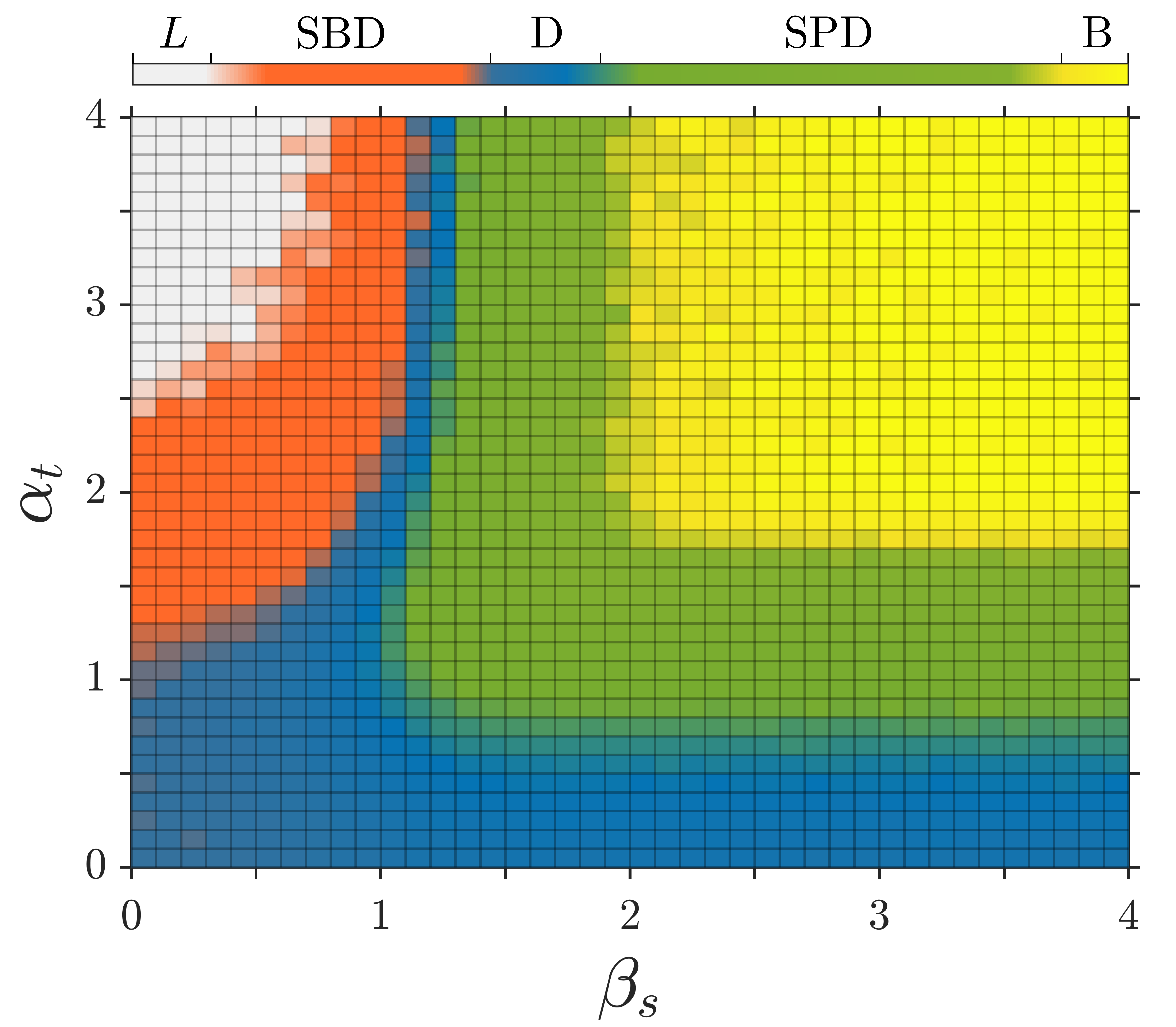}
\caption{Phase diagram with an overview of the dynamic behaviors of inhomogeneous DTQWs with the coin phases displaying different degrees of long-range space-time correlations, obtained from the exponent~$\gamma$: 
(L, gray) localized;
(SBD, orange) subdiffusive; 
(D,~blue) diffusive;
(SPD, green) superdiffusive; 
and (B, yellow) ballistic.}
%
%
%
%
%
%
%
\label{figure7}
\end{figure}

To end this section, we perform the complete mapping of the dynamics of inhomogeneous DTQWs with the coin phases displaying different degrees of long-range space-time correlations.
Figure~\ref{figure7} presents a phase diagram in the parameter space $\{ \alpha_t, \beta_s \}$, with the different labels and colors identifying the various dynamical regimes obtained from the exponent~$\gamma$: 
(L, gray) localized (large~$\alpha_t$ and small $\beta_s$);
(SBD, orange) subdiffusive (intermediate to large $\alpha_t$ and small $\beta_s$); 
(D,~blue) diffusive (small $\alpha_t$ and any $\beta_s$, and intermediate to large~$\alpha_t$ and intermediate $\beta_s$);
(SPD, green) superdiffusive (intermediate to large $\alpha_t$ and intermediate $\beta_s$, and intermediate $\alpha_t$ and intermediate to large $\beta_s$); 
and (B, yellow) ballistic (large~$\alpha_t$~and~$\beta_s$).

\section{Conclusions}

The importance of studying disorder effects on discrete-time quantum walks (DTQWs) can be hardly overstated. 
Actually, advances in the understanding of this issue can impact on practical implementations of associated qubit states, as it enables greater control over the system dynamics in the presence of environmental interactions. 

In this work, we have investigated the role of temporal and spatial random inhomogeneities in the phases of the quantum coin operator.
%
%
We have considered time and space correlations in the stochastic sequences of coin phases, described by a fractional Brownian motion with power-law spectrum, $S(k) \sim 1/k^\nu$, where the exponent $\nu$ drives the long-range character of temporal and spatial correlations, $\nu \to \alpha_t$ and $\nu \to \beta_s$, respectively. 

A suitable tunning of the degree of such correlations leads to several dynamic regimes in DTQWs. 
For instance, strong temporal correlations and weak spatial correlations give rise to an Anderson-like localized behavior of the quantum walker. 
Moreover, subdiffusive, diffusive, superdiffusive, and ballistic dynamics can be also found, e.g., with the latter arising in the presence of both spatial and temporal strong long-range correlations. 
We have presented a phase diagram with a mapping of these dynamical regimes in the $\{ \alpha_t, \beta_s \}$ parameter space. 

DTQWs may serve as a versatile platform to approach diverse phenomena, from quantum computation to cybersecutiry, to name a few. 
%
%
From these findings, we have shown that it is possible to control the degree of dynamic spreading of the qubit wave packet by properly adjusting the long-range correlation properties assigned to the inhomogeneous quantum coin operator. 
We hope our work can stimulate further theoretical and experimental research to advance the understanding of the interplay between disorder effects on inhomogeneous correlations and the resulting dynamics of DTQWs.
%


\section*{Acknowledgments}


This work was partially supported by the Brazilian agencies CNPq (Conselho Nacional de Desenvolvimento Cient\'{\i}fico e Tecnol\'ogico) and FACEPE (Funda\c{c}\~ao de Amparo a Ci\^encia e Tecnologia do Estado de Pernambuco).

\bibliography{references.bib}

\begin{thebibliography}{53}%
\makeatletter
\providecommand \@ifxundefined [1]{%
 \@ifx{#1\undefined}
}%
\providecommand \@ifnum [1]{%
 \ifnum #1\expandafter \@firstoftwo
 \else \expandafter \@secondoftwo
 \fi
}%
\providecommand \@ifx [1]{%
 \ifx #1\expandafter \@firstoftwo
 \else \expandafter \@secondoftwo
 \fi
}%
\providecommand \natexlab [1]{#1}%
\providecommand \enquote  [1]{``#1''}%
\providecommand \bibnamefont  [1]{#1}%
\providecommand \bibfnamefont [1]{#1}%
\providecommand \citenamefont [1]{#1}%
\providecommand \href@noop [0]{\@secondoftwo}%
\providecommand \href [0]{\begingroup \@sanitize@url \@href}%
\providecommand \@href[1]{\@@startlink{#1}\@@href}%
\providecommand \@@href[1]{\endgroup#1\@@endlink}%
\providecommand \@sanitize@url [0]{\catcode `\\12\catcode `\$12\catcode
  `\&12\catcode `\#12\catcode `\^12\catcode `\_12\catcode `\%12\relax}%
\providecommand \@@startlink[1]{}%
\providecommand \@@endlink[0]{}%
\providecommand \url  [0]{\begingroup\@sanitize@url \@url }%
\providecommand \@url [1]{\endgroup\@href {#1}{\urlprefix }}%
\providecommand \urlprefix  [0]{URL }%
\providecommand \Eprint [0]{\href }%
\providecommand \doibase [0]{https://doi.org/}%
\providecommand \selectlanguage [0]{\@gobble}%
\providecommand \bibinfo  [0]{\@secondoftwo}%
\providecommand \bibfield  [0]{\@secondoftwo}%
\providecommand \translation [1]{[#1]}%
\providecommand \BibitemOpen [0]{}%
\providecommand \bibitemStop [0]{}%
\providecommand \bibitemNoStop [0]{.\EOS\space}%
\providecommand \EOS [0]{\spacefactor3000\relax}%
\providecommand \BibitemShut  [1]{\csname bibitem#1\endcsname}%
\let\auto@bib@innerbib\@empty
\bibitem [{\citenamefont {Montanaro}(2016)}]{n1}%
  \BibitemOpen
  \bibfield  {author} {\bibinfo {author} {\bibfnamefont {A.}~\bibnamefont
  {Montanaro}},\ }\bibfield  {title} {\bibinfo {title} {Quantum algorithms: an
  overview},\ }\href {https://doi.org/10.1038/npjqi.2015.23} {\bibfield
  {journal} {\bibinfo  {journal} {npj Quantum Information}\ }\textbf {\bibinfo
  {volume} {2}},\ \bibinfo {pages} {1} (\bibinfo {year} {2016})}\BibitemShut
  {NoStop}%
\bibitem [{\citenamefont {Xia}\ \emph {et~al.}(2020)\citenamefont {Xia},
  \citenamefont {Liu}, \citenamefont {Nie}, \citenamefont {Fu}, \citenamefont
  {Wan},\ and\ \citenamefont {Kong}}]{n2}%
  \BibitemOpen
  \bibfield  {author} {\bibinfo {author} {\bibfnamefont {F.}~\bibnamefont
  {Xia}}, \bibinfo {author} {\bibfnamefont {J.}~\bibnamefont {Liu}}, \bibinfo
  {author} {\bibfnamefont {H.}~\bibnamefont {Nie}}, \bibinfo {author}
  {\bibfnamefont {Y.}~\bibnamefont {Fu}}, \bibinfo {author} {\bibfnamefont
  {L.}~\bibnamefont {Wan}},\ and\ \bibinfo {author} {\bibfnamefont
  {X.}~\bibnamefont {Kong}},\ }\bibfield  {title} {\bibinfo {title} {Random
  walks: A review of algorithms and applications},\ }\href
  {https://doi.org/10.1109/TETCI.2019.2952908} {\bibfield  {journal} {\bibinfo
  {journal} {IEEE Transactions on Emerging Topics in Computational
  Intelligence}\ }\textbf {\bibinfo {volume} {4}},\ \bibinfo {pages} {95}
  (\bibinfo {year} {2020})}\BibitemShut {NoStop}%
\bibitem [{\citenamefont {Venegas-Andraca}(2012)}]{venegas2012quantum}%
  \BibitemOpen
  \bibfield  {author} {\bibinfo {author} {\bibfnamefont {S.~E.}\ \bibnamefont
  {Venegas-Andraca}},\ }\bibfield  {title} {\bibinfo {title} {Quantum walks: a
  comprehensive review},\ }\href {https://doi.org/10.1007/s11128-012-0432-5}
  {\bibfield  {journal} {\bibinfo  {journal} {Quantum Information Processing}\
  }\textbf {\bibinfo {volume} {11}},\ \bibinfo {pages} {1015} (\bibinfo {year}
  {2012})}\BibitemShut {NoStop}%
\bibitem [{\citenamefont {Kempe}(2003)}]{kempe2003quantum}%
  \BibitemOpen
  \bibfield  {author} {\bibinfo {author} {\bibfnamefont {J.}~\bibnamefont
  {Kempe}},\ }\bibfield  {title} {\bibinfo {title} {Quantum random walks: an
  introductory overview},\ }\href
  {https://doi.org/10.1080/00107151031000110776} {\bibfield  {journal}
  {\bibinfo  {journal} {Contemporary Physics}\ }\textbf {\bibinfo {volume}
  {44}},\ \bibinfo {pages} {307} (\bibinfo {year} {2003})}\BibitemShut
  {NoStop}%
\bibitem [{\citenamefont {Aharonov}\ \emph {et~al.}(1993)\citenamefont
  {Aharonov}, \citenamefont {Davidovich},\ and\ \citenamefont
  {Zagury}}]{PhysRevA.48.1687}%
  \BibitemOpen
  \bibfield  {author} {\bibinfo {author} {\bibfnamefont {Y.}~\bibnamefont
  {Aharonov}}, \bibinfo {author} {\bibfnamefont {L.}~\bibnamefont
  {Davidovich}},\ and\ \bibinfo {author} {\bibfnamefont {N.}~\bibnamefont
  {Zagury}},\ }\bibfield  {title} {\bibinfo {title} {Quantum random walks},\
  }\href {https://doi.org/10.1103/PhysRevA.48.1687} {\bibfield  {journal}
  {\bibinfo  {journal} {Physical Review A}\ }\textbf {\bibinfo {volume} {48}},\
  \bibinfo {pages} {1687} (\bibinfo {year} {1993})}\BibitemShut {NoStop}%
\bibitem [{\citenamefont {Lovett}\ \emph {et~al.}(2010)\citenamefont {Lovett},
  \citenamefont {Cooper}, \citenamefont {Everitt}, \citenamefont {Trevers},\
  and\ \citenamefont {Kendon}}]{PhysRevA.81.042330}%
  \BibitemOpen
  \bibfield  {author} {\bibinfo {author} {\bibfnamefont {N.~B.}\ \bibnamefont
  {Lovett}}, \bibinfo {author} {\bibfnamefont {S.}~\bibnamefont {Cooper}},
  \bibinfo {author} {\bibfnamefont {M.}~\bibnamefont {Everitt}}, \bibinfo
  {author} {\bibfnamefont {M.}~\bibnamefont {Trevers}},\ and\ \bibinfo {author}
  {\bibfnamefont {V.}~\bibnamefont {Kendon}},\ }\bibfield  {title} {\bibinfo
  {title} {Universal quantum computation using the discrete-time quantum
  walk},\ }\href {https://doi.org/10.1103/PhysRevA.81.042330} {\bibfield
  {journal} {\bibinfo  {journal} {Physical Review A}\ }\textbf {\bibinfo
  {volume} {81}},\ \bibinfo {pages} {042330} (\bibinfo {year}
  {2010})}\BibitemShut {NoStop}%
\bibitem [{\citenamefont {Singh}\ \emph {et~al.}(2021)\citenamefont {Singh},
  \citenamefont {Chawla}, \citenamefont {Sarkar},\ and\ \citenamefont
  {Chandrashekar}}]{singh2021universal}%
  \BibitemOpen
  \bibfield  {author} {\bibinfo {author} {\bibfnamefont {S.}~\bibnamefont
  {Singh}}, \bibinfo {author} {\bibfnamefont {P.}~\bibnamefont {Chawla}},
  \bibinfo {author} {\bibfnamefont {A.}~\bibnamefont {Sarkar}},\ and\ \bibinfo
  {author} {\bibfnamefont {C.}~\bibnamefont {Chandrashekar}},\ }\bibfield
  {title} {\bibinfo {title} {Universal quantum computing using single-particle
  discrete-time quantum walk},\ }\href
  {https://doi.org/10.1038/s41598-021-91033-5} {\bibfield  {journal} {\bibinfo
  {journal} {Scientific Reports}\ }\textbf {\bibinfo {volume} {11}},\ \bibinfo
  {pages} {1} (\bibinfo {year} {2021})}\BibitemShut {NoStop}%
\bibitem [{\citenamefont {Shenvi}\ \emph {et~al.}(2003)\citenamefont {Shenvi},
  \citenamefont {Kempe},\ and\ \citenamefont {Whaley}}]{PhysRevA.67.052307}%
  \BibitemOpen
  \bibfield  {author} {\bibinfo {author} {\bibfnamefont {N.}~\bibnamefont
  {Shenvi}}, \bibinfo {author} {\bibfnamefont {J.}~\bibnamefont {Kempe}},\ and\
  \bibinfo {author} {\bibfnamefont {K.~B.}\ \bibnamefont {Whaley}},\ }\bibfield
   {title} {\bibinfo {title} {Quantum random-walk search algorithm},\ }\href
  {https://doi.org/10.1103/PhysRevA.67.052307} {\bibfield  {journal} {\bibinfo
  {journal} {Physical Review A}\ }\textbf {\bibinfo {volume} {67}},\ \bibinfo
  {pages} {052307} (\bibinfo {year} {2003})}\BibitemShut {NoStop}%
\bibitem [{\citenamefont {Portugal}(2013)}]{portugal2013quantum}%
  \BibitemOpen
  \bibfield  {author} {\bibinfo {author} {\bibfnamefont {R.}~\bibnamefont
  {Portugal}},\ }\href@noop {} {\emph {\bibinfo {title} {Quantum Walks and
  Search Algorithms}}}\ (\bibinfo  {publisher} {Springer, New York},\ \bibinfo
  {year} {2013})\BibitemShut {NoStop}%
\bibitem [{\citenamefont {T\'oth}(2012)}]{PhysRevA.85.022322}%
  \BibitemOpen
  \bibfield  {author} {\bibinfo {author} {\bibfnamefont {G.}~\bibnamefont
  {T\'oth}},\ }\bibfield  {title} {\bibinfo {title} {Multipartite entanglement
  and high-precision metrology},\ }\href
  {https://doi.org/10.1103/PhysRevA.85.022322} {\bibfield  {journal} {\bibinfo
  {journal} {Physical Review A}\ }\textbf {\bibinfo {volume} {85}},\ \bibinfo
  {pages} {022322} (\bibinfo {year} {2012})}\BibitemShut {NoStop}%
\bibitem [{\citenamefont {Abd El-Latif}\ \emph {et~al.}(2021)\citenamefont {Abd
  El-Latif}, \citenamefont {Abd-El-Atty}, \citenamefont {Mehmood},
  \citenamefont {Muhammad}, \citenamefont {Venegas-Andraca},\ and\
  \citenamefont {Peng}}]{abd2021quantum}%
  \BibitemOpen
  \bibfield  {author} {\bibinfo {author} {\bibfnamefont {A.~A.}\ \bibnamefont
  {Abd El-Latif}}, \bibinfo {author} {\bibfnamefont {B.}~\bibnamefont
  {Abd-El-Atty}}, \bibinfo {author} {\bibfnamefont {I.}~\bibnamefont
  {Mehmood}}, \bibinfo {author} {\bibfnamefont {K.}~\bibnamefont {Muhammad}},
  \bibinfo {author} {\bibfnamefont {S.~E.}\ \bibnamefont {Venegas-Andraca}},\
  and\ \bibinfo {author} {\bibfnamefont {J.}~\bibnamefont {Peng}},\ }\bibfield
  {title} {\bibinfo {title} {Quantum-inspired blockchain-based cybersecurity:
  securing smart edge utilities in iot-based smart cities},\ }\href
  {https://doi.org/10.1016/j.ipm.2021.102549} {\bibfield  {journal} {\bibinfo
  {journal} {Information Processing \& Management}\ }\textbf {\bibinfo {volume}
  {58}},\ \bibinfo {pages} {102549} (\bibinfo {year} {2021})}\BibitemShut
  {NoStop}%
\bibitem [{\citenamefont {Abd El-Latif}\ \emph {et~al.}(2020)\citenamefont {Abd
  El-Latif}, \citenamefont {Abd-El-Atty}, \citenamefont {Venegas-Andraca},
  \citenamefont {Elwahsh}, \citenamefont {Piran}, \citenamefont {Bashir},
  \citenamefont {Song},\ and\ \citenamefont {Mazurczyk}}]{abd2020providing}%
  \BibitemOpen
  \bibfield  {author} {\bibinfo {author} {\bibfnamefont {A.~A.}\ \bibnamefont
  {Abd El-Latif}}, \bibinfo {author} {\bibfnamefont {B.}~\bibnamefont
  {Abd-El-Atty}}, \bibinfo {author} {\bibfnamefont {S.~E.}\ \bibnamefont
  {Venegas-Andraca}}, \bibinfo {author} {\bibfnamefont {H.}~\bibnamefont
  {Elwahsh}}, \bibinfo {author} {\bibfnamefont {M.~J.}\ \bibnamefont {Piran}},
  \bibinfo {author} {\bibfnamefont {A.~K.}\ \bibnamefont {Bashir}}, \bibinfo
  {author} {\bibfnamefont {O.-Y.}\ \bibnamefont {Song}},\ and\ \bibinfo
  {author} {\bibfnamefont {W.}~\bibnamefont {Mazurczyk}},\ }\bibfield  {title}
  {\bibinfo {title} {Providing end-to-end security using quantum walks in iot
  networks},\ }\href {https://doi.org/10.1109/ACCESS.2020.2992820} {\bibfield
  {journal} {\bibinfo  {journal} {IEEE Access}\ }\textbf {\bibinfo {volume}
  {8}},\ \bibinfo {pages} {92687} (\bibinfo {year} {2020})}\BibitemShut
  {NoStop}%
\bibitem [{\citenamefont {Chandrashekar}(2006)}]{PhysRevA.74.032307}%
  \BibitemOpen
  \bibfield  {author} {\bibinfo {author} {\bibfnamefont {C.~M.}\ \bibnamefont
  {Chandrashekar}},\ }\bibfield  {title} {\bibinfo {title} {Implementing the
  one-dimensional quantum (hadamard) walk using a bose-einstein condensate},\
  }\href {https://doi.org/10.1103/PhysRevA.74.032307} {\bibfield  {journal}
  {\bibinfo  {journal} {Physical Review A}\ }\textbf {\bibinfo {volume} {74}},\
  \bibinfo {pages} {032307} (\bibinfo {year} {2006})}\BibitemShut {NoStop}%
\bibitem [{\citenamefont {Dadras}\ \emph {et~al.}(2018)\citenamefont {Dadras},
  \citenamefont {Gresch}, \citenamefont {Groiseau}, \citenamefont {Wimberger},\
  and\ \citenamefont {Summy}}]{PhysRevLett.121.070402}%
  \BibitemOpen
  \bibfield  {author} {\bibinfo {author} {\bibfnamefont {S.}~\bibnamefont
  {Dadras}}, \bibinfo {author} {\bibfnamefont {A.}~\bibnamefont {Gresch}},
  \bibinfo {author} {\bibfnamefont {C.}~\bibnamefont {Groiseau}}, \bibinfo
  {author} {\bibfnamefont {S.}~\bibnamefont {Wimberger}},\ and\ \bibinfo
  {author} {\bibfnamefont {G.~S.}\ \bibnamefont {Summy}},\ }\bibfield  {title}
  {\bibinfo {title} {Quantum walk in momentum space with a bose-einstein
  condensate},\ }\href {https://doi.org/10.1103/PhysRevLett.121.070402}
  {\bibfield  {journal} {\bibinfo  {journal} {Physical Review Letters}\
  }\textbf {\bibinfo {volume} {121}},\ \bibinfo {pages} {070402} (\bibinfo
  {year} {2018})}\BibitemShut {NoStop}%
\bibitem [{\citenamefont {Kitagawa}\ \emph {et~al.}(2010)\citenamefont
  {Kitagawa}, \citenamefont {Rudner}, \citenamefont {Berg},\ and\ \citenamefont
  {Demler}}]{PhysRevA.82.033429}%
  \BibitemOpen
  \bibfield  {author} {\bibinfo {author} {\bibfnamefont {T.}~\bibnamefont
  {Kitagawa}}, \bibinfo {author} {\bibfnamefont {M.~S.}\ \bibnamefont
  {Rudner}}, \bibinfo {author} {\bibfnamefont {E.}~\bibnamefont {Berg}},\ and\
  \bibinfo {author} {\bibfnamefont {E.}~\bibnamefont {Demler}},\ }\bibfield
  {title} {\bibinfo {title} {Exploring topological phases with quantum walks},\
  }\href {https://doi.org/10.1103/PhysRevA.82.033429} {\bibfield  {journal}
  {\bibinfo  {journal} {Physical Review A}\ }\textbf {\bibinfo {volume} {82}},\
  \bibinfo {pages} {033429} (\bibinfo {year} {2010})}\BibitemShut {NoStop}%
\bibitem [{\citenamefont {Barkhofen}\ \emph {et~al.}(2017)\citenamefont
  {Barkhofen}, \citenamefont {Nitsche}, \citenamefont {Elster}, \citenamefont
  {Lorz}, \citenamefont {G\'abris}, \citenamefont {Jex},\ and\ \citenamefont
  {Silberhorn}}]{PhysRevA.96.033846}%
  \BibitemOpen
  \bibfield  {author} {\bibinfo {author} {\bibfnamefont {S.}~\bibnamefont
  {Barkhofen}}, \bibinfo {author} {\bibfnamefont {T.}~\bibnamefont {Nitsche}},
  \bibinfo {author} {\bibfnamefont {F.}~\bibnamefont {Elster}}, \bibinfo
  {author} {\bibfnamefont {L.}~\bibnamefont {Lorz}}, \bibinfo {author}
  {\bibfnamefont {A.}~\bibnamefont {G\'abris}}, \bibinfo {author}
  {\bibfnamefont {I.}~\bibnamefont {Jex}},\ and\ \bibinfo {author}
  {\bibfnamefont {C.}~\bibnamefont {Silberhorn}},\ }\bibfield  {title}
  {\bibinfo {title} {Measuring topological invariants in disordered
  discrete-time quantum walks},\ }\href
  {https://doi.org/10.1103/PhysRevA.96.033846} {\bibfield  {journal} {\bibinfo
  {journal} {Physical Review A}\ }\textbf {\bibinfo {volume} {96}},\ \bibinfo
  {pages} {033846} (\bibinfo {year} {2017})}\BibitemShut {NoStop}%
\bibitem [{\citenamefont {Buarque}\ and\ \citenamefont
  {Dias}(2020)}]{PhysRevA.101.023802}%
  \BibitemOpen
  \bibfield  {author} {\bibinfo {author} {\bibfnamefont {A.~R.~C.}\
  \bibnamefont {Buarque}}\ and\ \bibinfo {author} {\bibfnamefont {W.~S.}\
  \bibnamefont {Dias}},\ }\bibfield  {title} {\bibinfo {title} {Self-trapped
  quantum walks},\ }\href {https://doi.org/10.1103/PhysRevA.101.023802}
  {\bibfield  {journal} {\bibinfo  {journal} {Physical Review A}\ }\textbf
  {\bibinfo {volume} {101}},\ \bibinfo {pages} {023802} (\bibinfo {year}
  {2020})}\BibitemShut {NoStop}%
\bibitem [{\citenamefont {Passos}\ and\ \citenamefont
  {Buarque}(2022)}]{PhysRevA.106.062407}%
  \BibitemOpen
  \bibfield  {author} {\bibinfo {author} {\bibfnamefont {F.~S.}\ \bibnamefont
  {Passos}}\ and\ \bibinfo {author} {\bibfnamefont {A.~R.~C.}\ \bibnamefont
  {Buarque}},\ }\bibfield  {title} {\bibinfo {title} {Nonlinear flip-flop
  quantum walks through potential barriers},\ }\href
  {https://doi.org/10.1103/PhysRevA.106.062407} {\bibfield  {journal} {\bibinfo
   {journal} {Physical Review A}\ }\textbf {\bibinfo {volume} {106}},\ \bibinfo
  {pages} {062407} (\bibinfo {year} {2022})}\BibitemShut {NoStop}%
\bibitem [{\citenamefont {Kendon}\ and\ \citenamefont
  {Tregenna}(2003)}]{PhysRevA.67.042315}%
  \BibitemOpen
  \bibfield  {author} {\bibinfo {author} {\bibfnamefont {V.}~\bibnamefont
  {Kendon}}\ and\ \bibinfo {author} {\bibfnamefont {B.}~\bibnamefont
  {Tregenna}},\ }\bibfield  {title} {\bibinfo {title} {Decoherence can be
  useful in quantum walks},\ }\href
  {https://doi.org/10.1103/PhysRevA.67.042315} {\bibfield  {journal} {\bibinfo
  {journal} {Physical Review A}\ }\textbf {\bibinfo {volume} {67}},\ \bibinfo
  {pages} {042315} (\bibinfo {year} {2003})}\BibitemShut {NoStop}%
\bibitem [{\citenamefont {Schreiber}\ \emph {et~al.}(2011)\citenamefont
  {Schreiber}, \citenamefont {Cassemiro}, \citenamefont
  {Poto\ifmmode~\check{c}\else \v{c}\fi{}ek}, \citenamefont {G\'abris},
  \citenamefont {Jex},\ and\ \citenamefont
  {Silberhorn}}]{PhysRevLett.106.180403}%
  \BibitemOpen
  \bibfield  {author} {\bibinfo {author} {\bibfnamefont {A.}~\bibnamefont
  {Schreiber}}, \bibinfo {author} {\bibfnamefont {K.~N.}\ \bibnamefont
  {Cassemiro}}, \bibinfo {author} {\bibfnamefont {V.}~\bibnamefont
  {Poto\ifmmode~\check{c}\else \v{c}\fi{}ek}}, \bibinfo {author} {\bibfnamefont
  {A.}~\bibnamefont {G\'abris}}, \bibinfo {author} {\bibfnamefont
  {I.}~\bibnamefont {Jex}},\ and\ \bibinfo {author} {\bibfnamefont
  {C.}~\bibnamefont {Silberhorn}},\ }\bibfield  {title} {\bibinfo {title}
  {Decoherence and disorder in quantum walks: From ballistic spread to
  localization},\ }\href {https://doi.org/10.1103/PhysRevLett.106.180403}
  {\bibfield  {journal} {\bibinfo  {journal} {Physical Review Letters}\
  }\textbf {\bibinfo {volume} {106}},\ \bibinfo {pages} {180403} (\bibinfo
  {year} {2011})}\BibitemShut {NoStop}%
\bibitem [{\citenamefont {Vieira}\ \emph {et~al.}(2013)\citenamefont {Vieira},
  \citenamefont {Amorim},\ and\ \citenamefont
  {Rigolin}}]{PhysRevLett.111.180503}%
  \BibitemOpen
  \bibfield  {author} {\bibinfo {author} {\bibfnamefont {R.}~\bibnamefont
  {Vieira}}, \bibinfo {author} {\bibfnamefont {E.~P.~M.}\ \bibnamefont
  {Amorim}},\ and\ \bibinfo {author} {\bibfnamefont {G.}~\bibnamefont
  {Rigolin}},\ }\bibfield  {title} {\bibinfo {title} {Dynamically disordered
  quantum walk as a maximal entanglement generator},\ }\href
  {https://doi.org/10.1103/PhysRevLett.111.180503} {\bibfield  {journal}
  {\bibinfo  {journal} {Physical Review Letters}\ }\textbf {\bibinfo {volume}
  {111}},\ \bibinfo {pages} {180503} (\bibinfo {year} {2013})}\BibitemShut
  {NoStop}%
\bibitem [{\citenamefont {Vakulchyk}\ \emph {et~al.}(2017)\citenamefont
  {Vakulchyk}, \citenamefont {Fistul}, \citenamefont {Qin},\ and\ \citenamefont
  {Flach}}]{PhysRevB.96.144204}%
  \BibitemOpen
  \bibfield  {author} {\bibinfo {author} {\bibfnamefont {I.}~\bibnamefont
  {Vakulchyk}}, \bibinfo {author} {\bibfnamefont {M.~V.}\ \bibnamefont
  {Fistul}}, \bibinfo {author} {\bibfnamefont {P.}~\bibnamefont {Qin}},\ and\
  \bibinfo {author} {\bibfnamefont {S.}~\bibnamefont {Flach}},\ }\bibfield
  {title} {\bibinfo {title} {Anderson localization in generalized discrete-time
  quantum walks},\ }\href {https://doi.org/10.1103/PhysRevB.96.144204}
  {\bibfield  {journal} {\bibinfo  {journal} {Physical Review B}\ }\textbf
  {\bibinfo {volume} {96}},\ \bibinfo {pages} {144204} (\bibinfo {year}
  {2017})}\BibitemShut {NoStop}%
\bibitem [{\citenamefont {Geraldi}\ \emph {et~al.}(2019)\citenamefont
  {Geraldi}, \citenamefont {Laneve}, \citenamefont {Bonavena}, \citenamefont
  {Sansoni}, \citenamefont {Ferraz}, \citenamefont {Fratalocchi}, \citenamefont
  {Sciarrino}, \citenamefont {Cuevas},\ and\ \citenamefont
  {Mataloni}}]{PhysRevLett.123.140501}%
  \BibitemOpen
  \bibfield  {author} {\bibinfo {author} {\bibfnamefont {A.}~\bibnamefont
  {Geraldi}}, \bibinfo {author} {\bibfnamefont {A.}~\bibnamefont {Laneve}},
  \bibinfo {author} {\bibfnamefont {L.~D.}\ \bibnamefont {Bonavena}}, \bibinfo
  {author} {\bibfnamefont {L.}~\bibnamefont {Sansoni}}, \bibinfo {author}
  {\bibfnamefont {J.}~\bibnamefont {Ferraz}}, \bibinfo {author} {\bibfnamefont
  {A.}~\bibnamefont {Fratalocchi}}, \bibinfo {author} {\bibfnamefont
  {F.}~\bibnamefont {Sciarrino}}, \bibinfo {author} {\bibfnamefont
  {A.}~\bibnamefont {Cuevas}},\ and\ \bibinfo {author} {\bibfnamefont
  {P.}~\bibnamefont {Mataloni}},\ }\bibfield  {title} {\bibinfo {title}
  {Experimental investigation of superdiffusion via coherent disordered quantum
  walks},\ }\href {https://doi.org/10.1103/PhysRevLett.123.140501} {\bibfield
  {journal} {\bibinfo  {journal} {Physical Review Letters}\ }\textbf {\bibinfo
  {volume} {123}},\ \bibinfo {pages} {140501} (\bibinfo {year}
  {2019})}\BibitemShut {NoStop}%
\bibitem [{\citenamefont {Geraldi}\ \emph {et~al.}(2021)\citenamefont
  {Geraldi}, \citenamefont {De}, \citenamefont {Laneve}, \citenamefont
  {Barkhofen}, \citenamefont {Sperling}, \citenamefont {Mataloni},\ and\
  \citenamefont {Silberhorn}}]{PhysRevResearch.3.023052}%
  \BibitemOpen
  \bibfield  {author} {\bibinfo {author} {\bibfnamefont {A.}~\bibnamefont
  {Geraldi}}, \bibinfo {author} {\bibfnamefont {S.}~\bibnamefont {De}},
  \bibinfo {author} {\bibfnamefont {A.}~\bibnamefont {Laneve}}, \bibinfo
  {author} {\bibfnamefont {S.}~\bibnamefont {Barkhofen}}, \bibinfo {author}
  {\bibfnamefont {J.}~\bibnamefont {Sperling}}, \bibinfo {author}
  {\bibfnamefont {P.}~\bibnamefont {Mataloni}},\ and\ \bibinfo {author}
  {\bibfnamefont {C.}~\bibnamefont {Silberhorn}},\ }\bibfield  {title}
  {\bibinfo {title} {Transient subdiffusion via disordered quantum walks},\
  }\href {https://doi.org/10.1103/PhysRevResearch.3.023052} {\bibfield
  {journal} {\bibinfo  {journal} {Physical Review Research}\ }\textbf {\bibinfo
  {volume} {3}},\ \bibinfo {pages} {023052} (\bibinfo {year}
  {2021})}\BibitemShut {NoStop}%
\bibitem [{\citenamefont {Schmitz}\ \emph {et~al.}(2009)\citenamefont
  {Schmitz}, \citenamefont {Matjeschk}, \citenamefont {Schneider},
  \citenamefont {Glueckert}, \citenamefont {Enderlein}, \citenamefont {Huber},\
  and\ \citenamefont {Schaetz}}]{PhysRevLett.103.090504}%
  \BibitemOpen
  \bibfield  {author} {\bibinfo {author} {\bibfnamefont {H.}~\bibnamefont
  {Schmitz}}, \bibinfo {author} {\bibfnamefont {R.}~\bibnamefont {Matjeschk}},
  \bibinfo {author} {\bibfnamefont {C.}~\bibnamefont {Schneider}}, \bibinfo
  {author} {\bibfnamefont {J.}~\bibnamefont {Glueckert}}, \bibinfo {author}
  {\bibfnamefont {M.}~\bibnamefont {Enderlein}}, \bibinfo {author}
  {\bibfnamefont {T.}~\bibnamefont {Huber}},\ and\ \bibinfo {author}
  {\bibfnamefont {T.}~\bibnamefont {Schaetz}},\ }\bibfield  {title} {\bibinfo
  {title} {Quantum walk of a trapped ion in phase space},\ }\href
  {https://doi.org/10.1103/PhysRevLett.103.090504} {\bibfield  {journal}
  {\bibinfo  {journal} {Physical Review Letters}\ }\textbf {\bibinfo {volume}
  {103}},\ \bibinfo {pages} {090504} (\bibinfo {year} {2009})}\BibitemShut
  {NoStop}%
\bibitem [{\citenamefont {Karski}\ \emph {et~al.}(2009)\citenamefont {Karski},
  \citenamefont {F{\"o}rster}, \citenamefont {Choi}, \citenamefont {Steffen},
  \citenamefont {Alt}, \citenamefont {Meschede},\ and\ \citenamefont
  {Widera}}]{karski2009quantum}%
  \BibitemOpen
  \bibfield  {author} {\bibinfo {author} {\bibfnamefont {M.}~\bibnamefont
  {Karski}}, \bibinfo {author} {\bibfnamefont {L.}~\bibnamefont {F{\"o}rster}},
  \bibinfo {author} {\bibfnamefont {J.-M.}\ \bibnamefont {Choi}}, \bibinfo
  {author} {\bibfnamefont {A.}~\bibnamefont {Steffen}}, \bibinfo {author}
  {\bibfnamefont {W.}~\bibnamefont {Alt}}, \bibinfo {author} {\bibfnamefont
  {D.}~\bibnamefont {Meschede}},\ and\ \bibinfo {author} {\bibfnamefont
  {A.}~\bibnamefont {Widera}},\ }\bibfield  {title} {\bibinfo {title} {Quantum
  walk in position space with single optically trapped atoms},\ }\href
  {https://doi.org/10.1126/science.1174436} {\bibfield  {journal} {\bibinfo
  {journal} {Science}\ }\textbf {\bibinfo {volume} {325}},\ \bibinfo {pages}
  {174} (\bibinfo {year} {2009})}\BibitemShut {NoStop}%
\bibitem [{\citenamefont {Perets}\ \emph {et~al.}(2008)\citenamefont {Perets},
  \citenamefont {Lahini}, \citenamefont {Pozzi}, \citenamefont {Sorel},
  \citenamefont {Morandotti},\ and\ \citenamefont
  {Silberberg}}]{PhysRevLett.100.170506}%
  \BibitemOpen
  \bibfield  {author} {\bibinfo {author} {\bibfnamefont {H.~B.}\ \bibnamefont
  {Perets}}, \bibinfo {author} {\bibfnamefont {Y.}~\bibnamefont {Lahini}},
  \bibinfo {author} {\bibfnamefont {F.}~\bibnamefont {Pozzi}}, \bibinfo
  {author} {\bibfnamefont {M.}~\bibnamefont {Sorel}}, \bibinfo {author}
  {\bibfnamefont {R.}~\bibnamefont {Morandotti}},\ and\ \bibinfo {author}
  {\bibfnamefont {Y.}~\bibnamefont {Silberberg}},\ }\bibfield  {title}
  {\bibinfo {title} {Realization of quantum walks with negligible decoherence
  in waveguide lattices},\ }\href
  {https://doi.org/10.1103/PhysRevLett.100.170506} {\bibfield  {journal}
  {\bibinfo  {journal} {Physical Review Letters}\ }\textbf {\bibinfo {volume}
  {100}},\ \bibinfo {pages} {170506} (\bibinfo {year} {2008})}\BibitemShut
  {NoStop}%
\bibitem [{\citenamefont {Zhang}\ \emph {et~al.}(2022)\citenamefont {Zhang},
  \citenamefont {Yang}, \citenamefont {Guo}, \citenamefont {Sun}, \citenamefont
  {Liu}, \citenamefont {Zhou}, \citenamefont {Xu}, \citenamefont {Xie},
  \citenamefont {Gong},\ and\ \citenamefont {Zhu}}]{PhysRevResearch.4.023042}%
  \BibitemOpen
  \bibfield  {author} {\bibinfo {author} {\bibfnamefont {R.}~\bibnamefont
  {Zhang}}, \bibinfo {author} {\bibfnamefont {R.}~\bibnamefont {Yang}},
  \bibinfo {author} {\bibfnamefont {J.}~\bibnamefont {Guo}}, \bibinfo {author}
  {\bibfnamefont {C.-W.}\ \bibnamefont {Sun}}, \bibinfo {author} {\bibfnamefont
  {Y.-C.}\ \bibnamefont {Liu}}, \bibinfo {author} {\bibfnamefont
  {H.}~\bibnamefont {Zhou}}, \bibinfo {author} {\bibfnamefont {P.}~\bibnamefont
  {Xu}}, \bibinfo {author} {\bibfnamefont {Z.}~\bibnamefont {Xie}}, \bibinfo
  {author} {\bibfnamefont {Y.-X.}\ \bibnamefont {Gong}},\ and\ \bibinfo
  {author} {\bibfnamefont {S.-N.}\ \bibnamefont {Zhu}},\ }\bibfield  {title}
  {\bibinfo {title} {Arbitrary coherent distributions in a programmable quantum
  walk},\ }\href {https://doi.org/10.1103/PhysRevResearch.4.023042} {\bibfield
  {journal} {\bibinfo  {journal} {Physical Review Research}\ }\textbf {\bibinfo
  {volume} {4}},\ \bibinfo {pages} {023042} (\bibinfo {year}
  {2022})}\BibitemShut {NoStop}%
\bibitem [{\citenamefont {Du}\ \emph {et~al.}(2003)\citenamefont {Du},
  \citenamefont {Li}, \citenamefont {Xu}, \citenamefont {Shi}, \citenamefont
  {Wu}, \citenamefont {Zhou},\ and\ \citenamefont {Han}}]{PhysRevA.67.042316}%
  \BibitemOpen
  \bibfield  {author} {\bibinfo {author} {\bibfnamefont {J.}~\bibnamefont
  {Du}}, \bibinfo {author} {\bibfnamefont {H.}~\bibnamefont {Li}}, \bibinfo
  {author} {\bibfnamefont {X.}~\bibnamefont {Xu}}, \bibinfo {author}
  {\bibfnamefont {M.}~\bibnamefont {Shi}}, \bibinfo {author} {\bibfnamefont
  {J.}~\bibnamefont {Wu}}, \bibinfo {author} {\bibfnamefont {X.}~\bibnamefont
  {Zhou}},\ and\ \bibinfo {author} {\bibfnamefont {R.}~\bibnamefont {Han}},\
  }\bibfield  {title} {\bibinfo {title} {Experimental implementation of the
  quantum random-walk algorithm},\ }\href
  {https://doi.org/10.1103/PhysRevA.67.042316} {\bibfield  {journal} {\bibinfo
  {journal} {Physical Review A}\ }\textbf {\bibinfo {volume} {67}},\ \bibinfo
  {pages} {042316} (\bibinfo {year} {2003})}\BibitemShut {NoStop}%
\bibitem [{\citenamefont {Ryan}\ \emph {et~al.}(2005)\citenamefont {Ryan},
  \citenamefont {Laforest}, \citenamefont {Boileau},\ and\ \citenamefont
  {Laflamme}}]{PhysRevA.72.062317}%
  \BibitemOpen
  \bibfield  {author} {\bibinfo {author} {\bibfnamefont {C.~A.}\ \bibnamefont
  {Ryan}}, \bibinfo {author} {\bibfnamefont {M.}~\bibnamefont {Laforest}},
  \bibinfo {author} {\bibfnamefont {J.~C.}\ \bibnamefont {Boileau}},\ and\
  \bibinfo {author} {\bibfnamefont {R.}~\bibnamefont {Laflamme}},\ }\bibfield
  {title} {\bibinfo {title} {Experimental implementation of a discrete-time
  quantum random walk on an nmr quantum-information processor},\ }\href
  {https://doi.org/10.1103/PhysRevA.72.062317} {\bibfield  {journal} {\bibinfo
  {journal} {Physical Review A}\ }\textbf {\bibinfo {volume} {72}},\ \bibinfo
  {pages} {062317} (\bibinfo {year} {2005})}\BibitemShut {NoStop}%
\bibitem [{\citenamefont {Flurin}\ \emph {et~al.}(2017)\citenamefont {Flurin},
  \citenamefont {Ramasesh}, \citenamefont {Hacohen-Gourgy}, \citenamefont
  {Martin}, \citenamefont {Yao},\ and\ \citenamefont
  {Siddiqi}}]{PhysRevX.7.031023}%
  \BibitemOpen
  \bibfield  {author} {\bibinfo {author} {\bibfnamefont {E.}~\bibnamefont
  {Flurin}}, \bibinfo {author} {\bibfnamefont {V.~V.}\ \bibnamefont
  {Ramasesh}}, \bibinfo {author} {\bibfnamefont {S.}~\bibnamefont
  {Hacohen-Gourgy}}, \bibinfo {author} {\bibfnamefont {L.~S.}\ \bibnamefont
  {Martin}}, \bibinfo {author} {\bibfnamefont {N.~Y.}\ \bibnamefont {Yao}},\
  and\ \bibinfo {author} {\bibfnamefont {I.}~\bibnamefont {Siddiqi}},\
  }\bibfield  {title} {\bibinfo {title} {Observing topological invariants using
  quantum walks in superconducting circuits},\ }\href
  {https://doi.org/10.1103/PhysRevX.7.031023} {\bibfield  {journal} {\bibinfo
  {journal} {Physical Review X}\ }\textbf {\bibinfo {volume} {7}},\ \bibinfo
  {pages} {031023} (\bibinfo {year} {2017})}\BibitemShut {NoStop}%
\bibitem [{\citenamefont {Wang}\ and\ \citenamefont {Manouchehri}(2013)}]{n3}%
  \BibitemOpen
  \bibfield  {author} {\bibinfo {author} {\bibfnamefont {J.}~\bibnamefont
  {Wang}}\ and\ \bibinfo {author} {\bibfnamefont {K.}~\bibnamefont
  {Manouchehri}},\ }\href@noop {} {\emph {\bibinfo {title} {Physical
  Implementation of Quantum Walks}}}\ (\bibinfo  {publisher} {Springer,
  Berlin},\ \bibinfo {year} {2013})\BibitemShut {NoStop}%
\bibitem [{\citenamefont {Shaji}\ and\ \citenamefont
  {Caves}(2007)}]{PhysRevA.76.032111}%
  \BibitemOpen
  \bibfield  {author} {\bibinfo {author} {\bibfnamefont {A.}~\bibnamefont
  {Shaji}}\ and\ \bibinfo {author} {\bibfnamefont {C.~M.}\ \bibnamefont
  {Caves}},\ }\bibfield  {title} {\bibinfo {title} {Qubit metrology and
  decoherence},\ }\href {https://doi.org/10.1103/PhysRevA.76.032111} {\bibfield
   {journal} {\bibinfo  {journal} {Physical Review A}\ }\textbf {\bibinfo
  {volume} {76}},\ \bibinfo {pages} {032111} (\bibinfo {year}
  {2007})}\BibitemShut {NoStop}%
\bibitem [{\citenamefont {Ahlbrecht}\ \emph {et~al.}(2012)\citenamefont
  {Ahlbrecht}, \citenamefont {Cedzich}, \citenamefont {Matjeschk},
  \citenamefont {Scholz}, \citenamefont {Werner},\ and\ \citenamefont
  {Werner}}]{ahlbrecht2012asymptotic}%
  \BibitemOpen
  \bibfield  {author} {\bibinfo {author} {\bibfnamefont {A.}~\bibnamefont
  {Ahlbrecht}}, \bibinfo {author} {\bibfnamefont {C.}~\bibnamefont {Cedzich}},
  \bibinfo {author} {\bibfnamefont {R.}~\bibnamefont {Matjeschk}}, \bibinfo
  {author} {\bibfnamefont {V.~B.}\ \bibnamefont {Scholz}}, \bibinfo {author}
  {\bibfnamefont {A.~H.}\ \bibnamefont {Werner}},\ and\ \bibinfo {author}
  {\bibfnamefont {R.~F.}\ \bibnamefont {Werner}},\ }\bibfield  {title}
  {\bibinfo {title} {Asymptotic behavior of quantum walks with spatio-temporal
  coin fluctuations},\ }\href {https://doi.org/10.1007/s11128-012-0389-4}
  {\bibfield  {journal} {\bibinfo  {journal} {Quantum Information Processing}\
  }\textbf {\bibinfo {volume} {11}},\ \bibinfo {pages} {1219} (\bibinfo {year}
  {2012})}\BibitemShut {NoStop}%
\bibitem [{\citenamefont {Vieira}\ \emph {et~al.}(2014)\citenamefont {Vieira},
  \citenamefont {Amorim},\ and\ \citenamefont {Rigolin}}]{PhysRevA.89.042307}%
  \BibitemOpen
  \bibfield  {author} {\bibinfo {author} {\bibfnamefont {R.}~\bibnamefont
  {Vieira}}, \bibinfo {author} {\bibfnamefont {E.~P.~M.}\ \bibnamefont
  {Amorim}},\ and\ \bibinfo {author} {\bibfnamefont {G.}~\bibnamefont
  {Rigolin}},\ }\bibfield  {title} {\bibinfo {title} {Entangling power of
  disordered quantum walks},\ }\href
  {https://doi.org/10.1103/PhysRevA.89.042307} {\bibfield  {journal} {\bibinfo
  {journal} {Physical Review A}\ }\textbf {\bibinfo {volume} {89}},\ \bibinfo
  {pages} {042307} (\bibinfo {year} {2014})}\BibitemShut {NoStop}%
\bibitem [{\citenamefont {Buarque}\ and\ \citenamefont
  {Dias}(2019)}]{PhysRevE.100.032106}%
  \BibitemOpen
  \bibfield  {author} {\bibinfo {author} {\bibfnamefont {A.~R.~C.}\
  \bibnamefont {Buarque}}\ and\ \bibinfo {author} {\bibfnamefont {W.~S.}\
  \bibnamefont {Dias}},\ }\bibfield  {title} {\bibinfo {title} {Aperiodic
  space-inhomogeneous quantum walks: Localization properties, energy spectra,
  and enhancement of entanglement},\ }\href
  {https://doi.org/10.1103/PhysRevE.100.032106} {\bibfield  {journal} {\bibinfo
   {journal} {Physical Review E}\ }\textbf {\bibinfo {volume} {100}},\ \bibinfo
  {pages} {032106} (\bibinfo {year} {2019})}\BibitemShut {NoStop}%
\bibitem [{\citenamefont {Wang}\ \emph {et~al.}(2018)\citenamefont {Wang},
  \citenamefont {Xu}, \citenamefont {Pan}, \citenamefont {Sun}, \citenamefont
  {Xu}, \citenamefont {Chen}, \citenamefont {Han}, \citenamefont {Li},\ and\
  \citenamefont {Guo}}]{wang2018dynamic}%
  \BibitemOpen
  \bibfield  {author} {\bibinfo {author} {\bibfnamefont {Q.-Q.}\ \bibnamefont
  {Wang}}, \bibinfo {author} {\bibfnamefont {X.-Y.}\ \bibnamefont {Xu}},
  \bibinfo {author} {\bibfnamefont {W.-W.}\ \bibnamefont {Pan}}, \bibinfo
  {author} {\bibfnamefont {K.}~\bibnamefont {Sun}}, \bibinfo {author}
  {\bibfnamefont {J.-S.}\ \bibnamefont {Xu}}, \bibinfo {author} {\bibfnamefont
  {G.}~\bibnamefont {Chen}}, \bibinfo {author} {\bibfnamefont {Y.-J.}\
  \bibnamefont {Han}}, \bibinfo {author} {\bibfnamefont {C.-F.}\ \bibnamefont
  {Li}},\ and\ \bibinfo {author} {\bibfnamefont {G.-C.}\ \bibnamefont {Guo}},\
  }\bibfield  {title} {\bibinfo {title} {Dynamic-disorder-induced enhancement
  of entanglement in photonic quantum walks},\ }\href
  {https://doi.org/10.1364/OPTICA.5.001136} {\bibfield  {journal} {\bibinfo
  {journal} {Optica}\ }\textbf {\bibinfo {volume} {5}},\ \bibinfo {pages}
  {1136} (\bibinfo {year} {2018})}\BibitemShut {NoStop}%
\bibitem [{\citenamefont {Zeng}\ and\ \citenamefont
  {Yong}(2017)}]{zeng2017discrete}%
  \BibitemOpen
  \bibfield  {author} {\bibinfo {author} {\bibfnamefont {M.}~\bibnamefont
  {Zeng}}\ and\ \bibinfo {author} {\bibfnamefont {E.~H.}\ \bibnamefont
  {Yong}},\ }\bibfield  {title} {\bibinfo {title} {Discrete-time quantum walk
  with phase disorder: localization and entanglement entropy},\ }\href
  {https://doi.org/10.1038/s41598-017-12077-0} {\bibfield  {journal} {\bibinfo
  {journal} {Scientific Reports}\ }\textbf {\bibinfo {volume} {7}},\ \bibinfo
  {pages} {1} (\bibinfo {year} {2017})}\BibitemShut {NoStop}%
\bibitem [{\citenamefont {Orthey}\ and\ \citenamefont {Amorim}(2019)}]{n5}%
  \BibitemOpen
  \bibfield  {author} {\bibinfo {author} {\bibfnamefont {A.~C.}\ \bibnamefont
  {Orthey}}\ and\ \bibinfo {author} {\bibfnamefont {E.~P.~M.}\ \bibnamefont
  {Amorim}},\ }\bibfield  {title} {\bibinfo {title} {Connecting velocity and
  entanglement in quantum walks},\ }\href
  {https://doi.org/10.1103/PhysRevA.99.032320} {\bibfield  {journal} {\bibinfo
  {journal} {Physical Review A}\ }\textbf {\bibinfo {volume} {99}},\ \bibinfo
  {pages} {032320} (\bibinfo {year} {2019})}\BibitemShut {NoStop}%
\bibitem [{\citenamefont {Mendes}\ \emph {et~al.}(2019)\citenamefont {Mendes},
  \citenamefont {Almeida}, \citenamefont {Lyra},\ and\ \citenamefont
  {de~Moura}}]{PhysRevE.99.022117}%
  \BibitemOpen
  \bibfield  {author} {\bibinfo {author} {\bibfnamefont {C.~V.~C.}\
  \bibnamefont {Mendes}}, \bibinfo {author} {\bibfnamefont {G.~M.~A.}\
  \bibnamefont {Almeida}}, \bibinfo {author} {\bibfnamefont {M.~L.}\
  \bibnamefont {Lyra}},\ and\ \bibinfo {author} {\bibfnamefont {F.~A. B.~F.}\
  \bibnamefont {de~Moura}},\ }\bibfield  {title} {\bibinfo {title}
  {Localization-delocalization transition in discrete-time quantum walks with
  long-range correlated disorder},\ }\href
  {https://doi.org/10.1103/PhysRevE.99.022117} {\bibfield  {journal} {\bibinfo
  {journal} {Physical Review E}\ }\textbf {\bibinfo {volume} {99}},\ \bibinfo
  {pages} {022117} (\bibinfo {year} {2019})}\BibitemShut {NoStop}%
\bibitem [{\citenamefont {Tregenna}\ \emph {et~al.}(2003)\citenamefont
  {Tregenna}, \citenamefont {Flanagan}, \citenamefont {Maile},\ and\
  \citenamefont {Kendon}}]{n4}%
  \BibitemOpen
  \bibfield  {author} {\bibinfo {author} {\bibfnamefont {B.}~\bibnamefont
  {Tregenna}}, \bibinfo {author} {\bibfnamefont {W.}~\bibnamefont {Flanagan}},
  \bibinfo {author} {\bibfnamefont {R.}~\bibnamefont {Maile}},\ and\ \bibinfo
  {author} {\bibfnamefont {V.}~\bibnamefont {Kendon}},\ }\bibfield  {title}
  {\bibinfo {title} {Controlling discrete quantum walks: coins and initial
  states},\ }\href {https://doi.org/10.1088/1367-2630/5/1/383} {\bibfield
  {journal} {\bibinfo  {journal} {New Journal of Physics}\ }\textbf {\bibinfo
  {volume} {5}},\ \bibinfo {pages} {83} (\bibinfo {year} {2003})}\BibitemShut
  {NoStop}%
\bibitem [{\citenamefont {de~Moura}\ and\ \citenamefont
  {Lyra}(1998)}]{mouralyra}%
  \BibitemOpen
  \bibfield  {author} {\bibinfo {author} {\bibfnamefont {F.~A. B.~F.}\
  \bibnamefont {de~Moura}}\ and\ \bibinfo {author} {\bibfnamefont {M.~L.}\
  \bibnamefont {Lyra}},\ }\bibfield  {title} {\bibinfo {title} {Delocalization
  in the 1d anderson model with long-range correlated disorder},\ }\href
  {https://doi.org/10.1103/PhysRevLett.81.3735} {\bibfield  {journal} {\bibinfo
   {journal} {Physical Review Letters}\ }\textbf {\bibinfo {volume} {81}},\
  \bibinfo {pages} {3735} (\bibinfo {year} {1998})}\BibitemShut {NoStop}%
\bibitem [{\citenamefont {Klafter}\ and\ \citenamefont {Sokolov}(2011)}]{nn1}%
  \BibitemOpen
  \bibfield  {author} {\bibinfo {author} {\bibfnamefont {J.}~\bibnamefont
  {Klafter}}\ and\ \bibinfo {author} {\bibfnamefont {I.~M.}\ \bibnamefont
  {Sokolov}},\ }\href@noop {} {\emph {\bibinfo {title} {First Steps in Random
  Walks}}}\ (\bibinfo  {publisher} {Oxford University Press, Oxford},\ \bibinfo
  {year} {2011})\BibitemShut {NoStop}%
\bibitem [{\citenamefont {Sassetti}\ \emph {et~al.}(1996)\citenamefont
  {Sassetti}, \citenamefont {Schomerus},\ and\ \citenamefont {Weiss}}]{nn2}%
  \BibitemOpen
  \bibfield  {author} {\bibinfo {author} {\bibfnamefont {M.}~\bibnamefont
  {Sassetti}}, \bibinfo {author} {\bibfnamefont {H.}~\bibnamefont
  {Schomerus}},\ and\ \bibinfo {author} {\bibfnamefont {U.}~\bibnamefont
  {Weiss}},\ }\bibfield  {title} {\bibinfo {title} {Subdiffusive and
  superdiffusive quantum transport and generalized duality},\ }\href
  {https://doi.org/10.1103/PhysRevB.53.R2914} {\bibfield  {journal} {\bibinfo
  {journal} {Physical Review B}\ }\textbf {\bibinfo {volume} {53}},\ \bibinfo
  {pages} {R2914} (\bibinfo {year} {1996})}\BibitemShut {NoStop}%
\bibitem [{\citenamefont {Bulchandani}\ and\ \citenamefont
  {Moore}(2020)}]{nn2b}%
  \BibitemOpen
  \bibfield  {author} {\bibinfo {author} {\bibfnamefont {K.~C.}\ \bibnamefont
  {Bulchandani}, \bibfnamefont {V.~B.}}\ and\ \bibinfo {author} {\bibfnamefont
  {J.~E.}\ \bibnamefont {Moore}},\ }\bibfield  {title} {\bibinfo {title}
  {Superdiffusive transport of energy in one-dimensional metals},\ }\href
  {https://doi.org/10.1073/pnas.191} {\bibfield  {journal} {\bibinfo  {journal}
  {Proceedings of the National Academy of Science U.S.A.}\ }\textbf {\bibinfo
  {volume} {117}},\ \bibinfo {pages} {12713} (\bibinfo {year}
  {2020})}\BibitemShut {NoStop}%
\bibitem [{\citenamefont {Grabarits}\ \emph {et~al.}(2022)\citenamefont
  {Grabarits}, \citenamefont {Kormos}, \citenamefont {Lovas},\ and\
  \citenamefont {Zar\'and}}]{nn2c}%
  \BibitemOpen
  \bibfield  {author} {\bibinfo {author} {\bibfnamefont {A.}~\bibnamefont
  {Grabarits}}, \bibinfo {author} {\bibfnamefont {M.}~\bibnamefont {Kormos}},
  \bibinfo {author} {\bibfnamefont {I.}~\bibnamefont {Lovas}},\ and\ \bibinfo
  {author} {\bibfnamefont {G.}~\bibnamefont {Zar\'and}},\ }\bibfield  {title}
  {\bibinfo {title} {Superdiffusive quantum work and adiabatic quantum
  evolution in finite temperature chaotic fermi systems},\ }\href
  {https://doi.org/10.1103/PhysRevB.106.064201} {\bibfield  {journal} {\bibinfo
   {journal} {Physical Review B}\ }\textbf {\bibinfo {volume} {106}},\ \bibinfo
  {pages} {064201} (\bibinfo {year} {2022})}\BibitemShut {NoStop}%
\bibitem [{\citenamefont {Barthelemy}\ and\ \citenamefont
  {Wiersma}(2008)}]{nn3}%
  \BibitemOpen
  \bibfield  {author} {\bibinfo {author} {\bibfnamefont {B.~J.}\ \bibnamefont
  {Barthelemy}, \bibfnamefont {P.}}\ and\ \bibinfo {author} {\bibfnamefont
  {D.}~\bibnamefont {Wiersma}},\ }\bibfield  {title} {\bibinfo {title} {A
  l\'evy flight for light},\ }\href
  {https://doi.org/https://doi.org/10.1038/nature06948} {\bibfield  {journal}
  {\bibinfo  {journal} {Nature (London)}\ }\textbf {\bibinfo {volume} {453}},\
  \bibinfo {pages} {495} (\bibinfo {year} {2008})}\BibitemShut {NoStop}%
\bibitem [{\citenamefont {Raposo}\ and\ \citenamefont {Gomes}(2015)}]{nn4}%
  \BibitemOpen
  \bibfield  {author} {\bibinfo {author} {\bibfnamefont {E.~P.}\ \bibnamefont
  {Raposo}}\ and\ \bibinfo {author} {\bibfnamefont {A.~S.~L.}\ \bibnamefont
  {Gomes}},\ }\bibfield  {title} {\bibinfo {title} {Analytical solution for the
  l\'evy-like steady-state distribution of intensities in random lasers},\
  }\href {https://doi.org/10.1103/PhysRevA.91.043827} {\bibfield  {journal}
  {\bibinfo  {journal} {Physical Review A}\ }\textbf {\bibinfo {volume} {91}},\
  \bibinfo {pages} {043827} (\bibinfo {year} {2015})}\BibitemShut {NoStop}%
\bibitem [{\citenamefont {Lima}\ \emph {et~al.}(2017)\citenamefont {Lima},
  \citenamefont {Pincheira}, \citenamefont {Raposo}, \citenamefont {Menezes},
  \citenamefont {de~Ara\'ujo}, \citenamefont {Gomes},\ and\ \citenamefont
  {Kashyap}}]{nn5}%
  \BibitemOpen
  \bibfield  {author} {\bibinfo {author} {\bibfnamefont {B.~C.}\ \bibnamefont
  {Lima}}, \bibinfo {author} {\bibfnamefont {P.~I.~R.}\ \bibnamefont
  {Pincheira}}, \bibinfo {author} {\bibfnamefont {E.~P.}\ \bibnamefont
  {Raposo}}, \bibinfo {author} {\bibfnamefont {L.~d.~S.}\ \bibnamefont
  {Menezes}}, \bibinfo {author} {\bibfnamefont {C.~B.}\ \bibnamefont
  {de~Ara\'ujo}}, \bibinfo {author} {\bibfnamefont {A.~S.~L.}\ \bibnamefont
  {Gomes}},\ and\ \bibinfo {author} {\bibfnamefont {R.}~\bibnamefont
  {Kashyap}},\ }\bibfield  {title} {\bibinfo {title} {Extreme-value statistics
  of intensities in a cw-pumped random fiber laser},\ }\href
  {https://doi.org/10.1103/PhysRevA.96.013834} {\bibfield  {journal} {\bibinfo
  {journal} {Physical Review A}\ }\textbf {\bibinfo {volume} {96}},\ \bibinfo
  {pages} {013834} (\bibinfo {year} {2017})}\BibitemShut {NoStop}%
\bibitem [{\citenamefont {de~Ara{\'u}jo}\ \emph {et~al.}(2017)\citenamefont
  {de~Ara{\'u}jo}, \citenamefont {Gomes},\ and\ \citenamefont {Raposo}}]{nn6}%
  \BibitemOpen
  \bibfield  {author} {\bibinfo {author} {\bibfnamefont {C.~B.}\ \bibnamefont
  {de~Ara{\'u}jo}}, \bibinfo {author} {\bibfnamefont {A.~S.~L.}\ \bibnamefont
  {Gomes}},\ and\ \bibinfo {author} {\bibfnamefont {E.~P.}\ \bibnamefont
  {Raposo}},\ }\bibfield  {title} {\bibinfo {title} {L{\'e}vy statistics and
  the glassy behavior of light in random fiber lasers},\ }\href
  {https://doi.org/10.3390/app7070644} {\bibfield  {journal} {\bibinfo
  {journal} {Applied Sciences}\ }\textbf {\bibinfo {volume} {7}},\ \bibinfo
  {pages} {644} (\bibinfo {year} {2017})}\BibitemShut {NoStop}%
\bibitem [{\citenamefont {Raposo}\ \emph {et~al.}(2003)\citenamefont {Raposo},
  \citenamefont {Buldyrev}, \citenamefont {da~Luz}, \citenamefont {Santos},
  \citenamefont {Stanley},\ and\ \citenamefont {Viswanathan}}]{nn7}%
  \BibitemOpen
  \bibfield  {author} {\bibinfo {author} {\bibfnamefont {E.~P.}\ \bibnamefont
  {Raposo}}, \bibinfo {author} {\bibfnamefont {S.~V.}\ \bibnamefont
  {Buldyrev}}, \bibinfo {author} {\bibfnamefont {M.~G.~E.}\ \bibnamefont
  {da~Luz}}, \bibinfo {author} {\bibfnamefont {M.~C.}\ \bibnamefont {Santos}},
  \bibinfo {author} {\bibfnamefont {H.~E.}\ \bibnamefont {Stanley}},\ and\
  \bibinfo {author} {\bibfnamefont {G.~M.}\ \bibnamefont {Viswanathan}},\
  }\bibfield  {title} {\bibinfo {title} {Dynamical robustness of l\'evy search
  strategies},\ }\href {https://doi.org/10.1103/PhysRevLett.91.240601}
  {\bibfield  {journal} {\bibinfo  {journal} {Physical Review Letters}\
  }\textbf {\bibinfo {volume} {91}},\ \bibinfo {pages} {240601} (\bibinfo
  {year} {2003})}\BibitemShut {NoStop}%
\bibitem [{\citenamefont {Wosniack}\ \emph {et~al.}(2015)\citenamefont
  {Wosniack}, \citenamefont {Santos}, \citenamefont {Raposo}, \citenamefont
  {Viswanathan},\ and\ \citenamefont {da~Luz}}]{nn8}%
  \BibitemOpen
  \bibfield  {author} {\bibinfo {author} {\bibfnamefont {M.~E.}\ \bibnamefont
  {Wosniack}}, \bibinfo {author} {\bibfnamefont {M.~C.}\ \bibnamefont
  {Santos}}, \bibinfo {author} {\bibfnamefont {E.~P.}\ \bibnamefont {Raposo}},
  \bibinfo {author} {\bibfnamefont {G.~M.}\ \bibnamefont {Viswanathan}},\ and\
  \bibinfo {author} {\bibfnamefont {M.~G.~E.}\ \bibnamefont {da~Luz}},\
  }\bibfield  {title} {\bibinfo {title} {Robustness of optimal random searches
  in fragmented environments},\ }\href
  {https://doi.org/10.1103/PhysRevE.91.052119} {\bibfield  {journal} {\bibinfo
  {journal} {Physical Review E}\ }\textbf {\bibinfo {volume} {91}},\ \bibinfo
  {pages} {052119} (\bibinfo {year} {2015})}\BibitemShut {NoStop}%
\bibitem [{\citenamefont {Santos}\ \emph {et~al.}(2005)\citenamefont {Santos},
  \citenamefont {Viswanathan}, \citenamefont {Raposo},\ and\ \citenamefont
  {da~Luz}}]{nn9}%
  \BibitemOpen
  \bibfield  {author} {\bibinfo {author} {\bibfnamefont {M.~C.}\ \bibnamefont
  {Santos}}, \bibinfo {author} {\bibfnamefont {G.~M.}\ \bibnamefont
  {Viswanathan}}, \bibinfo {author} {\bibfnamefont {E.~P.}\ \bibnamefont
  {Raposo}},\ and\ \bibinfo {author} {\bibfnamefont {M.~G.~E.}\ \bibnamefont
  {da~Luz}},\ }\bibfield  {title} {\bibinfo {title} {Optimization of random
  searches on regular lattices},\ }\href
  {https://doi.org/10.1103/PhysRevE.72.046143} {\bibfield  {journal} {\bibinfo
  {journal} {Physical Review E}\ }\textbf {\bibinfo {volume} {72}},\ \bibinfo
  {pages} {046143} (\bibinfo {year} {2005})}\BibitemShut {NoStop}%
\end{thebibliography}%

\end{document}